\documentclass[a4paper,11pt]{article}
\usepackage{jcappub}
\usepackage{amssymb}
\usepackage{graphicx}
\usepackage{amsmath}
\usepackage{hyperref}

\title{\boldmath Effective picture of bubble expansion}

\author[1,2]{Rong-Gen Cai,}
\author[1,3,4]{Shao-Jiang Wang$^*$}

\affiliation[1]{CAS Key Laboratory of Theoretical Physics, Institute of Theoretical Physics, Chinese Academy of Sciences, Beijing 100190, China}
\affiliation[2]{School of Fundamental Physics and Mathematical Sciences, Hangzhou Institute for Advanced Study (HIAS), University of Chinese Academy of Sciences, Hangzhou 310024, China}
\affiliation[3]{Quantum Universe Center and School of Physics, Korea Institute for Advanced Study (KIAS), Seoul 02455, Korea}
\affiliation[4]{Tufts Institute of Cosmology, Department of Physics and Astronomy, Tufts University, 574 Boston Avenue, Medford, Massachusetts 02155, USA}

\emailAdd{cairg@itp.ac.cn}
\emailAdd{schwang@cosmos.phy.tufts.edu (corresponding author)}

\abstract{Recently the thermal friction on an expanding bubble from the cosmic first-order phase transition has been calculated to all orders of the interactions between the bubble wall and thermal plasma, leading to a $\gamma^2$-scaling instead of the previously estimated $\gamma^1$-scaling for the thermal friction exerted on a fast-moving bubble wall with a Lorentz factor $\gamma$. We propose for the first time the effective equation of motion (EOM) for an expanding bubble wall in the presence of an arbitrary $\gamma$-scaling friction to compute the efficiency factor from bubble collisions, which, in the case of $\gamma^2$-scaling friction, is found to be larger than the recently updated estimation when the bubble walls collide after starting to approach a constant terminal velocity, leading to a slightly larger signal of the gravitational waves background from bubble collisions due to its quadratic dependence on the bubble collision efficiency factor, although the $\gamma^2$-scaling friction itself has already suppressed the contribution from bubble collisions compared to that with $\gamma^1$-scaling friction. We also suggest a phenomenological parameterization for the out-of-equilibrium term in the Boltzmann equation that could reproduce the recently found $(\gamma^2-1)$-scaling of the friction term in the effective EOM of an expanding bubble wall, which merits further study in future numerical simulations of bubble expansion and collisions.}

\begin{document}
\maketitle
\flushbottom

\section{Introduction}\label{sec:introduction}

The standard model (SM) of particle physics is incomplete at least in the sense of the puzzles of dark matter and baryon asymmetry of the Universe, of which the existing proposals could involve symmetry breakings of some beyond SM (BSM) physics that might produce cosmic bubbles during the first-order phase transitions (FOPT). The current and planned particle colliders might not be sufficient to probe such BSM physics, thus the resulted stochastic gravitational wave (GW) backgrounds \cite{Witten:1984rs,Hogan:1986qda} from FOPT  would be of particular interest \cite{Kosowsky:1991ua,Kosowsky:1992rz,Kosowsky:1992vn,Kamionkowski:1993fg,Huber:2008hg,Kosowsky:2001xp,Dolgov:2002ra,Nicolis:2003tg,Caprini:2006jb,Gogoberidze:2007an,Caprini:2009yp}, of which the reliable predictions \cite{Binetruy:2012ze,Caprini:2015zlo,Weir:2017wfa,Caprini:2019egz,Hindmarsh:2020hop} heavily rely on the dynamics of bubble nucleation \cite{Megevand:2016lpr,Jinno:2017ixd}, expansion \cite{Espinosa:2010hh,Ellis:2019oqb,Ellis:2020nnr,Giese:2020znk,Giese:2020rtr,Wang:2020nzm}, collision \cite{Caprini:2007xq,Caprini:2009fx,Weir:2016tov,Jinno:2016vai,Jinno:2017fby,Cutting:2018tjt,Jinno:2019bxw,Lewicki:2019gmv,Cutting:2020nla,Lewicki:2020jiv} and percolation \cite{Ellis:2018mja,Ellis:2020awk,Wang:2020jrd} as well as its ramification on the plasma motions \cite{Hindmarsh:2013xza,Hindmarsh:2015qta,Hindmarsh:2016lnk,Hindmarsh:2017gnf,Konstandin:2017sat,Niksa:2018ofa,Cutting:2019zws,Hindmarsh:2019phv,Jinno:2020eqg,Guo:2020grp}. Furthermore, the the presence of the first-order phase transition along with the understanding of bubble dynamics is also crucial for the realization of the electroweak baryogenesis \cite{Cohen:1990py,Cohen:1990it,Nelson:1991ab,Cohen:1994ss,Cohen:1993nk} and the primordial magnetic fields \cite{Hogan:1983zz,Quashnock:1988vs,Vachaspati:1991nm,Cheng:1994yr,Baym:1995fk}.

The bubble dynamics is well understood for the bubble nucleation in the vacuum background \cite{Coleman:1977py,Callan:1977pt} with the bubble wall velocity for the subsequent bubble expansion approaching the speed-of-light. However, the bubble nucleation in thermal background \cite{Linde:1980tt,Linde:1981zj} requires careful treatment \cite{Steinhardt:1981ct,Laine:1993ey} for the following bubble expansion due to the presence of thermal friction on the expanding bubble wall  \cite{Dine:1992wr,Liu:1992tn,Moore:1995ua,Moore:1995si} by the deviation from the thermal equilibrium across the bubble wall \cite{Turok:1992jp} governed by the Boltzmann equation \cite{Konstandin:2014zta}. Therefore, the direct calculation of bubble wall velocity is usually challenging and is only partially successful for some particular BSM models \cite{John:2000zq,Cline:2000nw,Carena:2000id,Carena:2002ss,Konstandin:2005cd,Cirigliano:2006dg,Kozaczuk:2015owa,Dorsch:2018pat,Friedlander:2020tnq}. Hence some phenomenological approaches \cite{Ignatius:1993qn,Moore:1995si,KurkiSuonio:1996rk} are adopted for the $\gamma$-scaling behavior of the friction term \cite{Megevand:2009ut,Megevand:2009gh,Espinosa:2010hh,Huber:2011aa,Megevand:2013hwa,Huber:2013kj} by evaluating the momentum transfer rate integrated across the bubble wall  \cite{Dine:1992wr,Liu:1992tn,Moore:1995ua,Moore:1995si} (see also \cite{Khlebnikov:1992bx,Arnold:1993wc} for the equivalent fluctuation-dissipation arguments), which is incorporated into an effective equation of motion (EOM) of the bubble wall expansion that is obtained by integrating the EOM  of the scalar field across the bubble wall \cite{Megevand:2009ut,Megevand:2009gh,Espinosa:2010hh,Huber:2011aa,Megevand:2013hwa,Huber:2013kj}. 

The estimation of the $\gamma$-scaling \footnote{For better wording, the phrase ``$\gamma$-scaling'' will refer to a generic scaling behavior while the phrase ``$\gamma^1$-scaling'' will refer to the specific choice of a linear $\gamma$-dependence. }  behavior of the friction term is full of turns and twists. It was first obtained in \cite{Bodeker:2009qy} that the leading-order (LO) friction caused by the changes of the effective mass during the $1\to1$ particle transmission and reflection in the vicinity of bubble wall is independent of the Lorentz factor $\gamma$ of the bubble wall velocity, 
\begin{align}\label{eq:pLO}
P_{1\to1}\approx\frac{\Delta m^2 T^2}{24}\equiv\Delta p_\mathrm{LO},
\end{align}
where $\Delta m^2\equiv\sum_ic_ig_i\Delta m_i^2$ sums over all particles ($c_i=1$ for bosons and $c_i=1/2$ for fermions) with their mass-square $\Delta m_i^2\equiv m_i^2(\phi_-)-m_i^2(\phi_+)$ changed across the bubble wall. This leads to a runaway expansion \cite{Espinosa:2010hh} since the LO friction is saturated at a finite value during expansion.
It was then realized in \cite{Bodeker:2017cim} that the next-to-leading-order (NLO) friction caused by $1\to2$ transition splitting of a fermion emitting a soft vector boson is proportional to $\gamma$, 
\begin{align}\label{eq:pNLO}
P_{1\to2}\approx\gamma g^2\Delta m_VT^3\equiv\gamma\Delta p_\mathrm{NLO},
\end{align}
where $g^2\Delta m_V=\sum_ig_i\lambda_i^2\Delta m_i$ sums over only gauge bosons with gauge couplings $\lambda_i$ and mass changes $\Delta m_i\equiv m_i(\phi_-)-m_i(\phi_+)$. This prevents the bubble wall from running away \cite{Ellis:2019oqb} since the NLO friction keeps growing until balancing the driving force eventually.
Recently, it was disputed in \cite{Hoeche:2020rsg} that re-summing multiple soft gauge bosons scattering to all orders reveals a $\gamma^2$-scaling friction \footnote{The subscript in $\Delta p_{N\mathrm{LO}}$ should not be confused with that in $\Delta p_\mathrm{NLO}$.}  \cite{Hoeche:2020rsg},
\begin{align}\label{eq:pNLO2}
P_{1\to N}\approx0.005\gamma^2g^2T^4\equiv\gamma^2\Delta p_{N\mathrm{LO}},
\end{align}
where $g^2=\sum_ig_i\lambda_i^2$ sums over all all gauge bosons to which the scalar field couples with coupling $\lambda_i$.  This leads to recently updated estimation \cite{Ellis:2020nnr}  for the bubble collision efficiency factor following the previous discussions \cite{Espinosa:2010hh,Ellis:2019oqb}. Nevertheless, the more general quantum field theoretic formula \cite{Mancha:2020fzw} found similar but not exactly the same $\gamma$-scaling behavior,
\begin{align}\label{eq:pNLO3}
\frac{F_\mathrm{fr}}{A}\equiv-\Delta p_\mathrm{fr}=(\gamma^2-1)T\Delta s,
\end{align}
where $\Delta s$ equals to the change in entropy density (of plasma) across the bubble wall if local thermal equilibrium is attained on the both sides of the bubble wall \cite{Balaji:2020yrx}.  Note that the $\gamma^2-1$ prefactor indicates the presence of the thermal friction only when the bubble wall is moving. However, the new result presented in \cite{Hoeche:2020rsg} was questioned in \cite{Vanvlasselaer:2020niz} for reproducing the correct limit in the vanishing masses of the fermion and the vector boson.

In this paper, we will take an open mind for the recent new result on the $\gamma^2$-scaling friction and re-calculate the bubble collision efficiency factor from an effective description on the bubble expansion. In section \ref{sec:review}, we review the scalar-plasma system and set up the conventions and notations for our later use; In section \ref{sec:EOM}, we reduce the scalar-plasma system into a wall-plasma system and derive an effective EOM for bubble wall expansion; In section \ref{sec:Lagrangian}, we propose an effective Lagrangian that could reproduce an arbitrary $\gamma$-scaling friction term in the effective EOM of bubble wall expansion; In section \ref{sec:efficiency}, we calculate the bubble collision efficiency factor, which turns out to be larger than the recently updated estimation under some circumstance. We conclude our discussion  in section \ref{sec:conclusion}.

\section{Scalar-plasma system : a review}\label{sec:review}

In this section, we will review the thermodynamics and relativistic hydrodynamics of scalar-plasma system to set the conventions and notations that we will use later on.

\subsection{Thermodynamics}\label{subsec:thermodynamics}

The background field method of quantum field theory at zero and a finite temperatures \cite{PhysRevD.9.1686,PhysRevD.9.3320} (see also \cite{Quiros:1999jp}) gives rise to the effective potential up to one-loop order as
\begin{align}
V_\mathrm{eff}(\phi)&=V_\mathrm{tree}(\phi)+V_\mathrm{1-loop}(\phi,T),\\
V_\mathrm{1-loop}(\phi,T)&=\sum\limits_{i=\mathrm{B,F}}\pm\frac12g_iT\sum\limits_{n=-\infty}^{+\infty}\int\frac{\mathrm{d}^3\vec{k}}{(2\pi)^3}\log\left[\vec{k}+\omega_n^2+m_i^2(\phi)\right],
\end{align}
where $g_i$ is the number of degrees of freedom for particle species $i$, the Matsubara frequencies $\omega_n=2n\pi T$ for bosons and $\omega_n=(2n+1)\pi T$ for fermions form the euclidean loop 4-momentum $k_E=(\omega_n,\vec{k})$, and the effective mass $m_i^2(\phi)=V''_\mathrm{tree}(\phi)$ is evaluated from the tree-level potential. The one-loop part of effective potential could be further split into the zero-temperature vacuum potential and the finite-temperature thermal potential,
\begin{align}
V_\mathrm{1-loop}(\phi,T)&=V_\mathrm{1-loop}^{T=0}(\phi)+V_\mathrm{1-loop}^{T\neq0}(\phi,T),\\
V_\mathrm{1-loop}^{T=0}(\phi)&=\sum\limits_{i=\mathrm{B,F}}\pm\frac12g_i\int\frac{\mathrm{d}^4k_E}{(2\pi)^4}\log[k_E^2+m_i^2(\phi)]\nonumber\\
&=\sum\limits_{i=\mathrm{B,F}}\pm g_i\frac{m_i^4(\phi)}{64\pi^2}\left[\log\frac{m_i^2(\phi)}{\mu^2}-C_i-C_\mathrm{UV}\right]\equiv V_\mathrm{CW}(\phi),\\
V_\mathrm{1-loop}^{T\neq0}(\phi,T)&=\sum\limits_{i=\mathrm{B,F}}\pm g_iT\int\frac{\mathrm{d}^3\vec{k}}{(2\pi)^3}\log\left[1\mp e^{-\frac{\sqrt{\vec{k}^2+m_i^2}}{T}}\right]\equiv T^4\sum_{i=\mathrm{B,F}}g_iJ_i\left(\frac{m_i^2}{T^2}\right)\label{eq:VT}
\end{align}
where the zero-temperature vacuum potential could be regularized as the Coleman-Weinberg potential in $4-\epsilon$ dimensions with $C_i=5/6\,(3/2)$ for gauge bosons (scalars and fermions) and $C_\mathrm{UV}\equiv\frac{2}{\epsilon}-\gamma_E+\log4\pi+\mathcal{O}(\epsilon)$. $\gamma_E$ is the  Euler constant. The dimensionless integrations for bosons/fermions in the finite-temperature thermal potential read
\begin{align}
J_{\mathrm{B/F}}(x)=\pm\int_0^\infty\frac{\mathrm{d}y}{2\pi^2}y^2\log\left(1\mp\exp(-\sqrt{x+y^2})\right),
\end{align}
which could be expanded at low temperature limit $m_i/T\gg1$ as
\begin{align}
J_\mathrm{B/F}\left(\frac{m_i^2}{T^2}\right)=-\left(\frac{m_i}{2\pi T}\right)^\frac32e^{-m_i/T}\left[1+\mathcal{O}\left(\frac{T}{m_i}\right)\right]
\end{align}
or at high temperature limit $m_i/T\ll1$ \cite{Quiros:1999jp} as
\begin{align}
J_\mathrm{B}\left(\frac{m_i^2}{T^2}\right)&=-\frac{\pi^2}{90}+\frac{1}{24}\left(\frac{m_i}{T}\right)^2-\frac{1}{12\pi}\left(\frac{m_i}{T}\right)^3-\frac{1}{32\pi^2}\left(\frac{m_i}{T}\right)^4\left(\log\frac{m_ie^{\gamma_E}}{4\pi T}-\frac34\right)\nonumber\\
&-\frac{1}{16\pi^{5/2}}\left(\frac{m_i}{T}\right)^4\sum\limits_{l=1}^{\infty}(-1)^l\frac{\zeta(2l+1)}{(l+1)!}\Gamma\left(l+\frac12\right)\left(\frac{m_i^2}{4\pi^2T^2}\right)^l
\end{align}
for bosons and 
\begin{align}
J_\mathrm{F}\left(\frac{m_i^2}{T^2}\right)&=-\frac78\frac{\pi^2}{90}+\frac{1}{48}\left(\frac{m_i}{T}\right)^2+\frac{1}{32\pi^2}\left(\frac{m_i}{T}\right)^4\left(\log\frac{m_ie^{\gamma_E}}{\pi T}-\frac34\right)\nonumber\\
&+\frac{1}{8\pi^{5/2}}\left(\frac{m_i}{T}\right)^4\sum\limits_{l=1}^{\infty}(-1)^l\frac{\zeta(2l+1)}{(l+1)!}\left(1-\frac{1}{2^{2l+1}}\right)\Gamma\left(l+\frac12\right)\left(\frac{m_i^2}{\pi^2T^2}\right)^l
\end{align}
for fermions. In particular, the finite-temperature thermal potential for massless particle species recovers the usual form of radiation energy density (hence the name \textit{scalar-plasma system}),
\begin{align}
V_\mathrm{1-loop}^{T\neq0, m_i=0}(\phi,T)=-\frac{\pi^2}{90}\left(\sum\limits_{i=\mathrm{B}}g_i+\frac78\sum\limits_{i=\mathrm{F}}g_i\right)T^4\equiv-\frac{\pi^2}{90}g_\mathrm{eff}T^4\equiv-\rho_\mathrm{R}.
\end{align}
For the sake of simplicity, we will just split the effective potential into temperature-independent and temperature-dependent parts as
\begin{align}
V_\mathrm{eff}(\phi,T)=V_0(\phi)+V_T(\phi,T).
\end{align}

The thermodynamics of scalar-plasma system is based on the identification of the total free energy density as the effective potential,
\begin{align}
\mathcal{F}(\phi,T)=V_\mathrm{eff}(\phi,T),
\end{align}
which directly leads to the corresponding definitions of pressure, entropy density, energy density, and enthalpy density of the scalar-plasma system as
\begin{align}
p&=-\mathcal{F}(\phi,T),\\
s&=\frac{\partial p}{\partial T}=-\frac{\partial\mathcal{F}}{\partial T},\\
e&=\mathcal{F}+Ts=\mathcal{F}-T\frac{\partial\mathcal{F}}{\partial T},\\
\omega&=e+p=Ts=T\frac{\partial p}{\partial T}=-T\frac{\partial\mathcal{F}}{\partial T},
\end{align}
respectively.  The whole scalar-plasma system is assumed to be approximately modeled as a perfect fluid system for simplicity with the total energy-momentum tensor
\begin{align}
T^{\mu\nu}=(e+p)u^\mu u^\nu+pg^{\mu\nu},
\end{align}
where $g^{\mu\nu}$ is usually chosen as the Minkowski metric $\eta^{\mu\nu}=\mathrm{diag}(-1,1,1,1)$ and the $u^\mu=\gamma(v)(1,\vec{v})$ is the four-velocity of a fluid element with its three-velocity $\vec{v}=\mathrm{d}\vec{x}/\mathrm{d}t$ evaluated at a position $\vec{x}$. The conservation of the total energy-momentum tensor $\partial_\mu T^{\mu\nu}=0$ is usually the starting point to derive the fluid velocity profile \cite{Ignatius:1993qn,KurkiSuonio:1996rk,Espinosa:2010hh} across the bubble wall with the bubble wall velocity as an input parameter, which, however, is not the focus of this paper. 

Recently in  \cite{Mancha:2020fzw} , the conservation equation $\partial_\mu T^{\mu\nu}=0$ in the vicinity of bubble wall is also used to derive the total pressure change across the bubble wall. To see this, note that the pressures from plasma and scalar in the bubble wall frame (with prime symbol), $p'_\mathrm{p}$ and $p'_\phi$, respectively, are the Lorentz boost of the those in the plasma frame (without prime symbol), $p_\mathrm{p}$ and $p_\phi$, respectively, namely,
\begin{align}
p'_\phi&=\gamma_\mathrm{w}^2(\rho_\phi+v_\mathrm{w}^2p_\phi)=(\gamma_\mathrm{w}^2-1)(\rho_\phi+p_\phi)+p_\phi=p_\phi,\\
p'_\mathrm{p}&=\gamma_\mathrm{w}^2(p_\mathrm{p}+v_\mathrm{w}^2\rho_\mathrm{p})=(\gamma_\mathrm{w}^2-1)(\rho_\mathrm{p}+p_\mathrm{p})+p_\mathrm{p}=(\gamma_\mathrm{w}^2-1)Ts+p_\mathrm{p},
\end{align}
where in the first line $\rho_\phi+p_\phi=0$ is used for the scalar sector in the true and false vacuums inside and outside of the bubble, respectively, and $Ts=(\rho_\phi+p_\phi)+(\rho_\mathrm{p}+p_\mathrm{p})=Ts_\mathrm{p}$ is used for the second line. Then the conservation equation at the interface of bubble wall with planar approximation implies the balance equation in the bubble wall frame as
\begin{align}
0=\Delta p'_\mathrm{p}+\Delta p'_\phi=\Delta(p'_\mathrm{p}+p'_\phi)=\Delta[(\gamma_\mathrm{w}^2-1)Ts+p_\mathrm{p}+p_\phi]=(\gamma_\mathrm{w}^2-1)T\Delta s+\Delta p,
\end{align}
thus the total pressure change reads
\begin{align}
\Delta p=-(\gamma_\mathrm{w}^2-1)T\Delta s,
\end{align}
which is present only when the bubble wall is moving ($\gamma_\mathrm{w}>1$), hence meets the expectation of the friction. This general argument assumes local thermal equilibrium across the moving bubble to maintain Lorentz symmetry.

\subsection{Hydrodynamics}\label{subsec:hydrodynamics}

We next turn to review the hydrodynamics closely following \cite{Hindmarsh:2020hop}. The total energy-momentum tensor could be divided into two parts contributed separately from the scalar field and thermal plasma,
\begin{align}
T_\phi^{\mu\nu}&=\partial^\mu\phi\partial^\nu\phi+\eta^{\mu\nu}\left[-\frac12(\partial\phi)^2-V_0(\phi)\right],\label{eq:Tphi}\\
T_\mathrm{p}^{\mu\nu}&=\sum\limits_{i=\mathrm{B,F}}g_i\int\frac{\mathrm{d}^3\vec{k}}{(2\pi)^3}\frac{k^\mu k^\nu}{k^0}\bigg|_{k^0=E_i(\vec{k})}f_i(x,k),\label{eq:Tp}
\end{align}
where $E_i(\vec{k})\equiv\sqrt{\vec{k}^2+m_i^2}$ is the energy of particle of species $i$ with momentum $\vec{k}$ and $f_i(x,k)$ is the distribution function counting the average number of particles of species $i$ with momentum $\vec{k}$ and energy $E_i(\vec{k})$ in a volume element $(\vec{x},\vec{x}+\mathrm{d}\vec{x})\times(\vec{k},\vec{k}+\mathrm{d}\vec{k})$ of the phase space at time $t=x^0$. Then the corresponding conservation equations of each energy-momentum tensor,
\begin{align}
\partial_\mu T_\phi^{\mu\nu}&\equiv(\square\phi-V'_0(\phi))\partial^\nu\phi=+f^\nu,\label{eq:dTphi}\\
\partial_\mu T_\mathrm{p}^{\mu\nu}&\equiv\sum\limits_{i=\mathrm{B,F}}g_i\int\frac{\mathrm{d}^3\vec{k}}{(2\pi)^3}\frac{k^\mu k^\nu}{E_i(\vec{k})}\partial_\mu f_i(x,k)=-f^\nu\label{eq:dTp}
\end{align}
should exhibit a flow of energy-momentum transfer between the interacting scalar and plasma sectors, which could be obtained from the relativistic Boltzmann equation,
\begin{align}\label{eq:Boltzmann}
\left(k^\mu\partial_\mu+m_iF_i^\mu\frac{\partial}{\partial k^\mu}\right)\Theta(k^0)\delta(k^2+m_i^2)f_i(x,k)=C[f_i],
\end{align}
as shown shortly below. 

\subsubsection*{Conservation equation with absence of external force}

Before that, it is worth noting that, with the absence of the external force $F_i^\mu=-\partial^\mu m_i$ that originated, for example, from a field-dependent particle mass by $F_i^\mu=-m'_i(\phi)\partial^\mu\phi$, the collision function $C[f_i]$ that vanishes identically,
\begin{align}
\int\frac{\mathrm{d}^3\vec{k}}{(2\pi)^32E_i(\vec{k})}\psi(x,k)C[f_i(x,k)]=0,
\end{align}
after integrated with the so-called collision invariant $\psi(x,k)=a(x)+b_\mu(x)k^\mu$ for arbitrary functions $a(x)$ and $b_\mu(x)$, implies the conservation law 
\begin{align}
0=\sum\limits_{i=\mathrm{B,F}}g_i\int\frac{\mathrm{d}^3\vec{k}}{(2\pi)^32E_i(\vec{k})}a(x)C[f_i]
=\frac12a(x)\sum\limits_{i=\mathrm{B,F}}g_i\int\frac{\mathrm{d}^3\vec{k}}{(2\pi)^3}\frac{k^\mu}{E_i(\vec{k})}\partial_\mu f_i
=\frac12a(x)\partial_\mu j^\mu
\end{align}
of the particle current
\begin{align}
j^\mu=\sum\limits_{i=\mathrm{B,F}}g_i\int\frac{\mathrm{d}^3\vec{k}}{(2\pi)^3}\frac{k^\mu}{E_i(\vec{k})}f_i(x,k)
\end{align}
for the choice $b_\mu(x)=0$ and the conservation law
\begin{align}
0=\sum\limits_{i=\mathrm{B,F}}g_i\int\frac{\mathrm{d}^3\vec{k}}{(2\pi)^32E_i(\vec{k})}b_\mu(x)k^\mu C[f_i]
=\frac12b_\mu(x)\sum\limits_{i=\mathrm{B,F}}g_i\int\frac{\mathrm{d}^3\vec{k}}{(2\pi)^3}\frac{k^\mu k^\nu}{E_i(\vec{k})}\partial_\nu f_i
=\frac12b_\mu(x)\partial_\nu T_\mathrm{p}^{\mu\nu}
\end{align}
of the energy-momentum tensor \eqref{eq:Tp}  for the choice $a(x)=0$. 

\subsubsection*{Conservation equation with presence of external force}

However, with the presence of the external force $F^\mu$, the energy-momentum transfer-flow $f^\mu$ between the scalar and plasma sectors could be derived from integrating the relativistic Boltzmann equation \eqref{eq:Boltzmann} as
\begin{align}
0&=\sum\limits_{i=\mathrm{B,F}}g_i\int\frac{\mathrm{d}^4k}{(2\pi)^4}k^\nu C[f_i]\\
&=\sum\limits_{i=\mathrm{B,F}}g_i\int\frac{\mathrm{d}^4k}{(2\pi)^4}k^\nu\left(k^\mu\partial_\mu+m_iF_i^\mu\frac{\partial}{\partial k^\mu}\right)\Theta(k^0)\delta(k^2+m_i^2)f_i\\
&=\partial_\mu\sum\limits_{i=\mathrm{B,F}}g_i\int\frac{\mathrm{d}^3\vec{k}}{(2\pi)^3}\frac{k^\mu k^\nu}{E_i(\vec{k})}f_i-\sum\limits_{i=\mathrm{B,F}}g_im_iF_i^\mu\frac{\partial k^\nu}{\partial k^\mu}\int\frac{\mathrm{d}^4k}{(2\pi)^4}\Theta(k^0)\delta(k^2+m_i^2)f_i\\
&=\frac12\partial_\mu T_\mathrm{p}^{\mu\nu}-\sum\limits_{i=\mathrm{B,F}}g_im_iF_i^\nu\int\frac{\mathrm{d}^3\vec{k}}{(2\pi)^3}\frac{f_i}{2E_i(\vec{k})},
\end{align}
where in the last two lines we have used $\partial k^\nu/\partial k^\mu=\delta^\nu_\mu$ after integration by part and 
\begin{align}
\int\frac{\mathrm{d}^4k}{(2\pi)^4}\Theta(k^0)\delta(k^2+m_i^2)=\int\frac{\mathrm{d}^3\vec{k}}{(2\pi)^3}\frac{1}{2E_i(\vec{k})}.
\end{align}
This directly leads to the form of the energy-momentum transfer-flow in the conservation equation \eqref{eq:dTp} as
\begin{align}\label{eq:dTpf}
\partial_\mu T_\mathrm{p}^{\mu\nu}=\sum\limits_{i=\mathrm{B,F}}2g_im_iF_i^\nu\int\frac{\mathrm{d}^3\vec{k}}{(2\pi)^3}\frac{f_i}{2E_i(\vec{k})}=-\partial^\nu\phi\sum\limits_{i=\mathrm{B,F}}g_i\frac{\mathrm{d}m_i^2}{\mathrm{d}\phi}\int\frac{\mathrm{d}^3\vec{k}}{(2\pi)^3}\frac{f_i}{2E_i(\vec{k})}\equiv-f^\nu.
\end{align}

\subsubsection*{Equations of motion for scalar and plasma}

If the distribution function could be further split into an equilibrium and a non-equilibrium parts, $f_i=f_i^\mathrm{eq}+\delta f_i$, with the equilibrium part of form
\begin{align}
f_i^\mathrm{eq}(x,k)=\frac{1}{e^{(E_i(\vec{k})-\mu)/T}\mp1},\quad i=\mathrm{B, F}
\end{align}
following the Bose-Einstein/Fermi-Dirac statistics for bosons/fermions, respectively, the energy-momentum transfer-flow for the equilibrium distribution function with negligible chemical potential $\mu\simeq0$ admits a simple connection to the temperature-dependent part of the effective potential \eqref{eq:VT} by
\begin{align}
\frac{\partial V_T}{\partial\phi}&=\sum\limits_{i=\mathrm{B,F}}\pm g_iT\int\frac{\mathrm{d}^3\vec{k}}{(2\pi)^3}\left(1\mp e^{-E_i(\vec{k})/T}\right)^{-1}(\mp)e^{-E_i(\vec{k})/T}\left(-\frac{1}{T}\right)\frac{\mathrm{d}E_i(\vec{k})}{\mathrm{d}\phi}\\
&=\sum\limits_{i=\mathrm{B,F}}g_i\int\frac{\mathrm{d}^3\vec{k}}{(2\pi)^3}\frac{\mathrm{d}E_i(\vec{k})}{\mathrm{d}\phi}f_i^\mathrm{eq}=\sum\limits_{i=\mathrm{B,F}}g_i\frac{\mathrm{d}m_i^2}{\mathrm{d}\phi}\int\frac{\mathrm{d}^3\vec{k}}{(2\pi)^3}\frac{f_i^\mathrm{eq}}{2E_i(\vec{k})},
\end{align}
which could be used to rewrite \eqref{eq:dTpf} as
\begin{align}\label{eq:dTpdf}
\partial_\mu T_\mathrm{p}^{\mu\nu}+\partial^\nu\phi\frac{\partial V_T}{\partial\phi}=-\partial^\nu\phi\sum\limits_{i=\mathrm{B,F}}g_i\frac{\mathrm{d}m_i^2}{\mathrm{d}\phi}\int\frac{\mathrm{d}^3\vec{k}}{(2\pi)^3}\frac{\delta f_i}{2E_i(\vec{k})}.
\end{align}
Similarly, the conservation equation \eqref{eq:dTphi} becomes
\begin{align}\label{eq:dTphidf}
\square\phi-\frac{\partial V_\mathrm{eff}}{\partial\phi}=\sum\limits_{i=\mathrm{B,F}}g_i\frac{\mathrm{d}m_i^2}{\mathrm{d}\phi}\int\frac{\mathrm{d}^3\vec{k}}{(2\pi)^3}\frac{\delta f_i}{2E_i(\vec{k})}.
\end{align}
The out-of-equilibrium term in the EOM \eqref{eq:dTpdf} and \eqref{eq:dTphidf} plays the role of thermal friction. 

\subsubsection*{Parameterization of the out-of-equilibrium term}

One of the popular parameterizations for the out-of-equilibrium term reads
\begin{align}\label{eq:parafriction1}
\sum\limits_{i=\mathrm{B,F}}g_i\frac{\mathrm{d}m_i^2}{\mathrm{d}\phi}\int\frac{\mathrm{d}^3\vec{k}}{(2\pi)^3}\frac{\delta f_i}{2E_i(\vec{k})}\equiv\eta_Tu^\mu\partial_\mu\phi,
\end{align}
where $\eta_T$ is some function of the scalar field and temperature as well as the fluid velocity. Note that although the friction term $\eta_T u^\mu\partial_\mu\phi$ is introduced in a Lorentz invariant form, the scalar-plasma system actually breaks the Lorentz invariance under the boosts. The first argument to support such parameterization is its relation to the conservation equation of entropy current of plasma. To see this, we have to first assume that the energy-momentum tensor of the plasma could also be modeled as a perfect fluid 
\footnote{
at least locally reasonable for a comoving observer with $u^\mu=(1,0,0,0)$ if we define the energy density and pressure of the plasma by
\begin{align}
e_\mathrm{p}&=\sum\limits_{i=\mathrm{B,F}}g_i\int\frac{\mathrm{d}^3\vec{k}}{(2\pi)^3}k^0f_i,\\
p_\mathrm{p}&=\sum\limits_{i=\mathrm{B,F}}g_i\int\frac{\mathrm{d}^3\vec{k}}{(2\pi)^3}\frac{\vec{k}^2}{k^0}f_i,
\end{align}
respectively, so as to reproduce \eqref{eq:Tp},
\begin{align}
T_\mathrm{p}^{\mu\nu}=\sum\limits_{i=\mathrm{B,F}}g_i\int\frac{\mathrm{d}^3\vec{k}}{(2\pi)^3}\left[\left(k^0+\frac{\vec{k}^2}{k^0}\right)u^\mu u^\nu+\eta^{\mu\nu}\frac{\vec{k}^2}{k^0}\right]f_i=\sum\limits_{i=\mathrm{B,F}}g_i\int\frac{\mathrm{d}^3\vec{k}}{(2\pi)^3}\frac{k^\mu k^\nu}{k^0}f_i.
\end{align}
},
\begin{align}
T_\mathrm{p}^{\mu\nu}=(e_\mathrm{p}+p_\mathrm{p})u^\mu u^\nu+p_\mathrm{p}\eta^{\mu\nu},
\end{align}
then multiplying \eqref{eq:dTpdf} with $u_\nu$ after using the perfect fluid ansatz, 
\begin{align}
u_\nu\partial_\mu[(e_\mathrm{p}+p_\mathrm{p})u^\mu u^\nu+p_\mathrm{p}\eta^{\mu\nu}]+\frac{\partial V_T}{\partial\phi}(u_\nu\partial^\nu\phi)=-\eta_T(u_\nu\partial^\nu\phi)(u^\mu\partial_\mu\phi),
\end{align}
followed by adopting $u_\nu u^\nu=-1$ and $u_\nu\partial_\mu u^\nu=0$, we would find the conservation equation for the plasma entropy current $S_\mathrm{p}^\mu$ defined by $(e_\mathrm{p}+p_\mathrm{p})u^\mu\equiv\omega_\mathrm{p}u^\mu\equiv Ts_\mathrm{p}u^\mu\equiv TS_\mathrm{p}^\mu$ as
\begin{align}
\partial_\mu S_\mathrm{p}^\mu=\frac{\eta_T}{T}(u\cdot\partial\phi)^2,
\end{align}
provided that $p_\mathrm{p}=-V_T(\phi,T)$.
Therefore, the conservation of entropy current of plasma is violated by the deviation of distribution function away from equilibrium. The second argument to choose such parameterization is that, when integrating the scalar EOM \eqref{eq:dTphidf} across the bubble wall as shown in section \ref{subsec:EOMf}, the resulted effective EOM of the bubble wall expansion could reproduce the $\gamma^1$-scaling friction term \eqref{eq:pNLO} that was originally estimated by evaluating the momentum transfer rate integrated across the bubble wall. 

However, with $\gamma^2$-scaling friction, such a parameterization for the out-of-equilibrium term should be modified in order to numerically simulate the bubble expansion in the thermal plasma. Perhaps a more crucial question to ask is how to pick one specific parameterization for the out-of-equilibrium term that is more appropriate than the others. One way to do that is to compare the numerical simulation results from the scalar EOM \eqref{eq:dTphidf} with some parameterization for the out-of-equilibrium term to the analytic results from an effective EOM of the bubble wall expansion as derived below in the next section.

\section{Wall-plasma system : effective EOM}\label{sec:EOM}

The time evolution of bubble expansion requires the full machinery of solving the combined Boltzmann equations \eqref{eq:dTpdf} and \eqref{eq:dTphidf}, leading to the solutions for the scalar field $\phi(t,\vec{x})$ and for the velocity profile of a fluid element $u^\mu(t,\vec{x})$ in the plasma. This scalar-plasma system could be further reduced into a wall-plasma system if we assume a field profile $\phi(t,r)=\phi(\gamma_\mathrm{w}(t)[r-r_\mathrm{w}(t)])\equiv\phi(r')$ satisfying $\phi(r'=-\infty)=\phi_-$ and $\phi(r'=+\infty)=\phi_+$ by measuring $r$, $r_\mathrm{w}$ and $r'$ with the bubble wall width at thin-wall limit. Here the bubble wall frame $(t', r')$ is boosted from the coordinates $(t,r)$ in the bubble center frame by the Lorentz transformation,
\begin{align}
r'&=\gamma_\mathrm{w}(t)[r-r_\mathrm{w}(t)],\\
t'&=\gamma_\mathrm{w}(t)[t-v_\mathrm{w}(t)r],
\end{align}
where we have assumed a spherical profile for the bubble expansion and $r_\mathrm{w}(t)$ is the wall position moving with velocity $v_\mathrm{w}(t)=\dot{r}_\mathrm{w}(t)$ and corresponding Lorentz factor $\gamma_\mathrm{w}(t)=1/\sqrt{1-v_\mathrm{w}^2(t)}$. Then the scalar EOM of the scalar-plasma system could be reduced into an effective EOM of the expanding bubble wall as shown below.

\subsection{Thermal bubble in thermal equilibrium}\label{subsec:EOM}

In the absence of the out-of-equilibrium friction term,  the EOM of scalar field with spherical profile reads
\begin{align}\label{eq:SEOM}
\nabla_\mu\nabla^\mu\phi
\equiv\frac{\partial^2\phi}{\partial r^2}+\frac{2}{r}\frac{\partial\phi}{\partial r}-\frac{\partial^2\phi}{\partial t^2}
=\frac{\partial V_\mathrm{eff}}{\partial\phi}.
\end{align}
We first converts above partial derivative terms written in the bubble center frame into bubble wall frame,
\begin{align}
\frac{\partial\phi}{\partial r}
&=\frac{\mathrm{d}\phi}{\mathrm{d}r'}\frac{\partial r'}{\partial r}=\phi'(r')\gamma_\mathrm{w},
\end{align}
\begin{align}
\frac{\partial^2\phi}{\partial r^2}
&=\frac{\mathrm{d}^2\phi}{\mathrm{d}r'^2}\left(\frac{\partial r'}{\partial r}\right)^2+\frac{\mathrm{d}\phi}{\mathrm{d}r'}\frac{\partial^2r'}{\partial r^2}=\phi''(r')\gamma_\mathrm{w}^2,
\end{align}
\begin{align}
\frac{\partial\phi}{\partial t}
&=\frac{\mathrm{d}\phi}{\mathrm{d}r'}\frac{\partial r'}{\partial t}
=\phi'(r')[\dot{\gamma}_\mathrm{w}(r-r_\mathrm{w})-\gamma_\mathrm{w}v_\mathrm{w}]
=\phi'(r')[(\dot{\gamma}_\mathrm{w}/\gamma_\mathrm{w})r'-\gamma_\mathrm{w}v_\mathrm{w}]\nonumber\\
&=\phi'(r')(\gamma_\mathrm{w}^2v_\mathrm{w}\ddot{r}_\mathrm{w}r'-\gamma_\mathrm{w}v_\mathrm{w})=\phi'(r')\gamma_\mathrm{w}v_\mathrm{w}(\gamma_\mathrm{w}\ddot{r}_\mathrm{w}r'-1),
\end{align}
\begin{align}
\frac{\partial^2\phi}{\partial t^2}
&=\frac{\mathrm{d}^2\phi}{\mathrm{d}r'^2}\left(\frac{\partial r'}{\partial t}\right)^2+\frac{\mathrm{d}\phi}{\mathrm{d}r'}\frac{\partial^2r'}{\partial t^2}\nonumber\\
&=\phi''(r')[\dot{\gamma}_\mathrm{w}(r-r_\mathrm{w})-\gamma_\mathrm{w}v_\mathrm{w}]^2+\phi'(r')[\ddot{\gamma}_\mathrm{w}(r-r_\mathrm{w})-2\dot{\gamma}_\mathrm{w}v_\mathrm{w}-\gamma_\mathrm{w}\ddot{r}_\mathrm{w}]\nonumber\\
&=\phi''(r')\gamma_\mathrm{w}^2v_\mathrm{w}^2(\gamma_\mathrm{w}\ddot{r}_\mathrm{w}r'-1)^2+\phi'(r')[(\ddot{\gamma}_\mathrm{w}/\gamma_\mathrm{w})r'-\gamma_\mathrm{w}(2\gamma_\mathrm{w}^2-1)\ddot{r}_\mathrm{w}],
\end{align}
where $\dot{\gamma}_\mathrm{w}=\gamma_\mathrm{w}^3v_\mathrm{w}\ddot{r}_\mathrm{w}$ and $\gamma_\mathrm{w}^2v_\mathrm{w}^2=\gamma_\mathrm{w}^2-1$ are used to simplify the expressions. Then, we define the bubble tension as
\begin{align}
\sigma=\int_{-\infty}^{+\infty}\mathrm{d}r' \phi'(r')^2,
\end{align}
which could be used to define the averaged value of some quantity $F$ by
\begin{align}
\langle F\rangle=\frac{1}{\sigma}\int_{-\infty}^{+\infty}\mathrm{d}r' \phi'(r')^2F(r').
\end{align}
Note that $\langle F\rangle=0$ if $F(r')$ is an odd function, for example, $\langle r'\rangle=0$. Next, we integrate the EOM \eqref{eq:SEOM} over $r'$ after multiplied by $\phi'(r')$, of which each terms read
\begin{align}
\int_{-\infty}^{+\infty}\mathrm{d}r'\phi'(r')\frac{\partial^2\phi}{\partial r^2}&=\int\mathrm{d}\left(\frac{\phi'^2}{2}\right)\gamma_\mathrm{w}^2=\frac{\gamma_\mathrm{w}^2}{2}\phi'(r')^2\bigg|_{-\infty}^{+\infty}=0,
\end{align}
\begin{align}
\int_{-\infty}^{+\infty}\mathrm{d}r'\phi'(r')\frac{2}{r}\frac{\partial\phi}{\partial r}&=\int_{-\infty}^{+\infty}\mathrm{d}r'\phi'(r')\frac{2}{r}\phi'(r')\gamma_\mathrm{w}\simeq\frac{2\sigma\gamma_\mathrm{w}}{r_\mathrm{w}},
\end{align}
\begin{align}
\int_{-\infty}^{+\infty}\mathrm{d}r'\phi'(r')\frac{\partial^2\phi}{\partial t^2}
&=\int_{-\infty}^{+\infty}\mathrm{d}\left(\frac{\phi'^2}{2}\right)\gamma_\mathrm{w}^2v_\mathrm{w}^2(\gamma_\mathrm{w}\ddot{r}_\mathrm{w}r'-1)^2\nonumber\\
&\quad+\int_{-\infty}^{+\infty}\mathrm{d}r'\phi'(r')^2[(\ddot{\gamma}_\mathrm{w}/\gamma_\mathrm{w})r'-\gamma_\mathrm{w}(2\gamma_\mathrm{w}^2-1)\ddot{r}_\mathrm{w}]\nonumber\\
&=\frac12\phi'(r')^2\gamma_\mathrm{w}^2v_\mathrm{w}^2(\gamma_\mathrm{w}\ddot{r}_\mathrm{w}r'-1)^2\bigg|_{-\infty}^{+\infty}\nonumber\\
&\quad-\int_{-\infty}^{+\infty}\phi'(r')^2\gamma_\mathrm{w}^2v_\mathrm{w}^2(\gamma_\mathrm{w}\ddot{r}_\mathrm{w}r'-1)\gamma_\mathrm{w}\ddot{r}_\mathrm{w}\mathrm{d}r'\nonumber\\
&\quad+\int_{-\infty}^{+\infty}\mathrm{d}r'\phi'(r')^2[(\ddot{\gamma}_\mathrm{w}/\gamma_\mathrm{w})r'-\gamma_\mathrm{w}(2\gamma_\mathrm{w}^2-1)\ddot{r}_\mathrm{w}]\nonumber\\
&=0+\sigma\gamma_\mathrm{w}^3v_\mathrm{w}^2\ddot{r}_\mathrm{w}-\sigma\gamma_\mathrm{w}(2\gamma_\mathrm{w}^2-1)\ddot{r}_\mathrm{w}=-\sigma\gamma_\mathrm{w}^3\ddot{r}_\mathrm{w},
\end{align}
\begin{align}
\int_{-\infty}^{+\infty}\mathrm{d}r'\phi'(r')\frac{\partial V_\mathrm{eff}}{\partial \phi}
&=\int_{\phi_-}^{\phi_+}\mathrm{d}\phi\left(\frac{\partial V_0}{\partial\phi}+\frac{\partial V_T}{\partial\phi}\right)\nonumber\\
&=\int_{\phi_-}^{\phi_+}\mathrm{d}\phi\frac{\mathrm{d}V_0}{\mathrm{d}\phi}+\sum\limits_{i=\mathrm{B,F}}g_iT^4\int_{\phi_-}^{\phi_+}\mathrm{d}\phi\,J'_i\left(\frac{m_i^2}{T^2}\right)\frac{1}{T^2}\frac{\mathrm{d}m_i^2}{\mathrm{d}\phi}\nonumber\\
&=V_0(\phi)\bigg|_{\phi_-}^{\phi_+}+\sum\limits_{i=\mathrm{B,F}}g_iT^4J_i\left(\frac{m_i^2(\phi)}{T^2}\right)\bigg|_{\phi_-}^{\phi_+}\nonumber\\
&\equiv-\Delta V_0-\Delta V_T=-\Delta V_\mathrm{eff}\equiv V_\mathrm{eff}(\phi_+)-V_\mathrm{eff}(\phi_-),
\end{align}
Finally, putting everything together directly leads to the effective equation of bubble expansion in thermal equilibrium,
\begin{align}
\sigma\gamma_\mathrm{w}^3\ddot{r}_\mathrm{w}+\frac{2\sigma\gamma_\mathrm{w}}{r_\mathrm{w}}=\Delta p_\mathrm{dr}
\end{align}
namely,
\begin{align}\label{eq:rw}
\ddot{r}_\mathrm{w}+2\frac{1-\dot{r}_\mathrm{w}^2}{r_\mathrm{w}}=\frac{\Delta p_\mathrm{dr}}{\sigma}\left(1-\dot{r}_\mathrm{w}^2\right)^\frac32,
\end{align}
where  $\Delta p_\mathrm{dr}\equiv-\Delta\mathcal{F}\equiv-\Delta V_\mathrm{eff}\equiv V_\mathrm{eff}(\phi_+)-V_\mathrm{eff}(\phi_-)$. 

\subsection{Thermal bubble out of thermal equilibrium}\label{subsec:EOMf}

In the presence of the parameterization  \eqref{eq:parafriction1} for the out-of-equilibrium term,  the EOM of scalar field with spherical profile reads
\begin{align}\label{eq:SEOMf}
\nabla_\mu\nabla^\mu\phi
\equiv\frac{\partial^2\phi}{\partial r^2}+\frac{2}{r}\frac{\partial\phi}{\partial r}-\frac{\partial^2\phi}{\partial t^2}
=\frac{\partial V_\mathrm{eff}}{\partial\phi}+\eta_T u^\mu\partial_\mu\phi,
\end{align}
of which the integration over $r'$ after multiplied by $\phi'(r')$,
\begin{align}\label{eq:EOMWall}
\sigma\gamma_\mathrm{w}^3\ddot{r}_\mathrm{w}+\frac{2\sigma\gamma_\mathrm{w}}{r_\mathrm{w}}=\Delta p_\mathrm{dr}+\Delta p_\mathrm{fr},
\end{align}
receives an extra contribution from the thermal friction of form
\begin{align}
\Delta p_\mathrm{fr}&=\int_{-\infty}^{+\infty}\mathrm{d}r'\phi'(r')\eta_T u^\mu\partial_\mu\phi\nonumber\\
&=\int_{-\infty}^{+\infty}\mathrm{d}r'\phi'(r')\eta_T\left(\gamma_\mathrm{p}\frac{\partial\phi}{\partial t}+\gamma_\mathrm{p}v_\mathrm{p}\frac{\partial \phi}{\partial r}\right)\nonumber\\
&=\int_{-\infty}^{+\infty}\mathrm{d}r'\phi'(r')^2\eta_T[\gamma_\mathrm{p}\gamma_\mathrm{w}v_\mathrm{w}(\gamma_\mathrm{w}\ddot{r}_\mathrm{w}r'-1)+\gamma_\mathrm{p}v_\mathrm{p}\gamma_\mathrm{w}]\nonumber\\
&=-\int_{-\infty}^{+\infty}\mathrm{d}r'\phi'(r')^2\eta_T\gamma_\mathrm{p}\gamma_\mathrm{w}(v_\mathrm{w}-v_\mathrm{p}).
\end{align}
This cannot be determined without first specifying the form of $\eta_T(\phi, T, v_\mathrm{p})$ and the relation between the plasma velocity profile $v_\mathrm{p}$ with bubble wall velocity $v_\mathrm{w}$. To roughly estimate the scaling of such a parameterization of the out-of-equilibrium term, we could look at the detonation case with an ultra-relativistic wall velocity where the fluid element of the plasma moves with the opposite velocity with respect to the bubble wall frame, $v'_\mathrm{p}=-v_\mathrm{w}$, so that the plasma velocity is static in the plasma frame, $v_\mathrm{p}=0$ and $\gamma_\mathrm{p}=1$, hence the friction term should scale as
\begin{align}
\Delta p_\mathrm{fr}=-\gamma_\mathrm{w}v_\mathrm{w}\sigma\langle\eta_T\rangle\propto-\gamma_\mathrm{w}v_\mathrm{w},
\end{align}
reproducing the $\gamma^1$-scaling behavior \eqref{eq:pNLO} in the ultra-relativistic limit $v_\mathrm{w}\approx1$.

This however does not meet the recently findings of $\gamma^2$-scaling or  $(\gamma^2-1)$-scaling for the friction force, which requires for another parameterization for the out-of-equilibrium term. We propose here for future study a parameterization of the out-of-equilibrium term,
\begin{align}
\sum\limits_{i=\mathrm{B,F}}g_i\frac{\mathrm{d}m_i^2}{\mathrm{d}\phi}\int\frac{\mathrm{d}^3\vec{k}}{(2\pi)^3}\frac{\delta f_i}{2E_i(\vec{k})}\equiv-\eta_T(u^\mu\partial_\mu\phi)^2,
\end{align}
which reproduces the recently estimated $\gamma^2$-scaling or  $(\gamma^2-1)$-scaling for the friction force,
\begin{align}
\Delta p_\mathrm{fr}&=-\int_{-\infty}^{+\infty}\mathrm{d}r'\phi'(r')^2\eta_T\gamma_\mathrm{w}^2v_\mathrm{w}^2(\gamma_\mathrm{w}\ddot{r}_\mathrm{w}r'-1)^2\nonumber\\
&=-\int_{-\infty}^{+\infty}\mathrm{d}r'\phi'(r')^2\eta_T\gamma_\mathrm{w}^2v_\mathrm{w}^2(1+\gamma_\mathrm{w}^2\ddot{r}_\mathrm{w}^2r'^2)\nonumber\\
&=-\gamma_\mathrm{w}^2v_\mathrm{w}^2\sigma\langle\eta_T\rangle\propto-(\gamma_\mathrm{w}^2-1).\label{eq:newfriction}
\end{align}
Here in the first line we have assumed again the detonation case  $v_\mathrm{p}=0$ and $\gamma_\mathrm{p}=1$, and in the second line we get rid of the odd term in $r'$, and in the last line we have focused on the late-time scaling when the bubble wall stops accelerating $\ddot{r}_\mathrm{w}\approx0$.

\section{Wall-plasma system :  effective Lagrangian}\label{sec:Lagrangian}

The forementioned parameterization for the out-of-equilibrium term is not easy to use in the effective EOM \eqref{eq:EOMWall}  since it usually requires the preknowledge of the fluid velocity $u^\mu$, which itself should be solved from an equation governing the fluid velocity profile of the thermal plasma \cite{Espinosa:2010hh} with the bubble wall velocity as an  input in the first place. Therefore, we might as well directly parameterize the friction term in the effective EOM \eqref{eq:EOMWall}  as 
\begin{align}
\Delta p_\mathrm{fr}=-\Delta p_\mathrm{LO}-h(\gamma)\Delta p_{N\mathrm{LO}},
\end{align}
which has been separated into the $\gamma$-independent and $\gamma$-dependent (with arbitrary $\gamma$-scaling function $h(\gamma)$) terms.  
However, with presence of a $\gamma$-dependent contribution to the friction term $\Delta p_\mathrm{fr}$, the effective EOM \eqref{eq:EOMWall}  actually does not respect the conservation law of the total energy of an expanding bubble. As we will see in section \ref{subsec:new}, the effective EOM for an expanding bubble wall in thermal plasma with $\gamma$-dependent friction is derived for the first time as 
\begin{align}\label{eq:EOMWallgamma}
\left(\sigma+\frac{r_\mathrm{w}}{3}\frac{\mathrm{d}|\Delta p_\mathrm{fr}|}{\mathrm{d}\gamma_\mathrm{w}}\right)\gamma_\mathrm{w}^3\ddot{r}_\mathrm{w}+\frac{2\sigma\gamma_\mathrm{w}}{r_\mathrm{w}}=\Delta p_\mathrm{dr}+\Delta p_\mathrm{fr},
\end{align}
which reduces to \eqref{eq:EOMWall} when the friction term is $\gamma$-independent. It is easy to check that the solution of  \eqref{eq:EOMWallgamma} obeys the conservation law of the total energy of an expanding bubble,
\begin{align}
E=4\pi\sigma r_\mathrm{w}^2\gamma_\mathrm{w}-\frac43\pi r_\mathrm{w}^3(\Delta p_\mathrm{dr}+\Delta p_\mathrm{fr}).
\end{align}
Note that the usual out-of-equilibrium calculations on the bubble wall velocity from the microscopic particle physics would give rise to some specific forms for the $\gamma$-scaling function $h(\gamma)$, which is however assumed to be generic in this paper to set up the general  framework for the calculation of the efficiency factor from bubble collisions.

\subsection{Thermal bubble in thermal equilibrium}

The effective equation \eqref{eq:rw} for bubble expansion in thermal equilibrium could be derived from an effective Lagrangian of form
\begin{align}
L=-4\pi\sigma R^2\sqrt{1-\dot{R}^2}+\frac43\pi R^3\Delta p_\mathrm{dr}
\end{align}
where the change of driving pressure $\Delta p_\mathrm{dr}\equiv-\Delta V_\mathrm{eff}\equiv V_\mathrm{eff}(\phi_+)-V_\mathrm{eff}(\phi_-)$, and $R(t)$ is just a simplified notation hereafter for the previously used $r_\mathrm{w}(t)$. The resulting Euler-Lagrangian equation
\begin{align}\label{eq:REOM0}
\ddot{R}+2\frac{1-\dot{R}^2}{R}=\frac{\Delta p_\mathrm{dr}}{\sigma}\left(1-\dot{R}^2\right)^\frac32,
\end{align}
has a simple form in terms of the Lorentz factor $\gamma=1/\sqrt{1-\dot{R}^2}$, namely,
\begin{align}\label{eq:GEOM0}
\frac{\mathrm{d}\gamma}{\mathrm{d}R}=\ddot{R}\gamma^3=\frac{\Delta p_\mathrm{dr}}{\sigma}-\frac{2\gamma}{R},
\end{align}
which could be analytically solved as
\begin{align}
\gamma(R)=\frac{\Delta p_\mathrm{dr}}{3\sigma}R+\frac{C}{R^2}
\end{align}
with $C$ fixed by the initial condition $\gamma(R_0)=1$. Here the initially static bubble radius  $R_0=2\sigma/\Delta p_\mathrm{dr}$ could be determined from
\begin{align}
\left.\frac{\mathrm{d}E}{\mathrm{d}R}\right|_{\gamma(t=0)=1}=0
\end{align}
by the conservation law of bubble energy $E=(\partial L/\partial\dot{R})\dot{R}-L=4\pi\sigma R^2\gamma-\frac43\pi R^3\Delta p_\mathrm{dr}$. The corresponding solution
\begin{align}\label{eq:gammaR0}
\gamma(R)=\frac{R_0^2}{R^2}+\frac{\Delta p_\mathrm{dr}}{\sigma}\frac{R^3-R_0^3}{3R^2}.
\end{align}
could be further simplified by normalizing  $R$ with respect to $R_0$ as
\begin{align}\label{eq:gamma0}
\gamma(R)=\frac23R+\frac{1}{3R^2},
\end{align}
where $R_0$ has been set to 1 for convenience. This gives rise to the familiar picture of bubble expansion that the bubble velocity
\begin{align}
\dot{R}(t)=\sqrt{1-\frac{9R^4}{(1+2R^3)^2}}
\end{align}
quickly approaches the speed-of-light right after nucleation if the driving pressure $\Delta p_\mathrm{dr}>0$. It is worth noting that, the effective equation  \eqref{eq:REOM0} or \eqref{eq:GEOM0} also applies to the case of bubble expansion in vacuum by replacing the driving pressure purely from its vacuum part as $\Delta p_\mathrm{dr}=-\Delta V_0\equiv V_0(\phi_+)-V_0(\phi_-)$. Nevertheless, the bubble expansion in thermal equilibrium can never be the realistic case since the thermal equilibrium should always be broken near the bubble wall, hence there comes the presence of the friction term in the effective EOM from the out-of-equilibrium term in Boltzmann equation as we discuss below.

\subsection{Thermal bubble out of thermal equilibrium}

To include the effect of a $\gamma$-dependent friction into the effective EOM \eqref{eq:REOM0}, we first review the previous approximated treatment and then introduce our effective description.

\subsubsection{Previous approximated estimation}\label{subsec:old}

Take the $\gamma^1$-scaling friction as an example, the previous approximated treatment  \cite{Ellis:2019oqb} simply inserts the NLO friction term by hand directly into the EOM \eqref{eq:REOM0}, 
\begin{align}\label{eq:REOM1}
\ddot{R}+2\frac{1-\dot{R}^2}{R}=\frac{\Delta p_\mathrm{dr}-\Delta p_\mathrm{LO}-\gamma\Delta p_\mathrm{NLO}}{\sigma}\left(1-\dot{R}^2\right)^\frac32,
\end{align}
or rewritten in terms of the Lorentz factor,
\begin{align}\label{eq:GEOM1}
\frac{\mathrm{d}\gamma}{\mathrm{d}R}+\frac{2\gamma}{R}=\frac{\Delta p_\mathrm{dr}-\Delta p_\mathrm{LO}-\gamma\Delta p_\mathrm{NLO}}{\sigma}\equiv\eta(\gamma_\mathrm{eq}-\gamma),
\end{align}
where abbreviations 
\begin{align}
\eta\equiv\frac{\Delta p_\mathrm{NLO}}{\sigma},\quad \gamma_\mathrm{eq}\equiv\frac{\Delta p_\mathrm{dr}-\Delta p_\mathrm{LO}}{\Delta p_\mathrm{NLO}}
\end{align}
are introduced for convenient parameterization. Directly solving \eqref{eq:GEOM1} gives rise to 
\begin{align}
\gamma(R)=C\frac{\mathrm{e}^{-\eta R}}{R^2}+\gamma_\mathrm{eq}\frac{\eta^2R^2-2\eta R+2}{\eta^2R^2},
\end{align}
which, after fixing the constant $C$ by the initial condition $\gamma(R_0)=1$, becomes
\begin{align}
\gamma(R)=\frac{\eta^2R_0^2-\gamma_\mathrm{eq}(2-2\eta R_0+\eta^2R_0^2)}{\eta^2R^2}\mathrm{e}^{-\eta(R-R_0)}+\frac{\gamma_\mathrm{eq}(2-2\eta R+\eta^2R^2)}{\eta^2R^2},
\end{align}
namely,
\begin{align}\label{eq:gammaeom1}
\gamma(R)=-\frac{(\gamma_\mathrm{eq}-1)^2(\gamma_\mathrm{eq}-2)}{2R^2}\mathrm{e}^{-\frac{2(R-1)}{\gamma_\mathrm{eq}-1}}+\frac{2\gamma_\mathrm{eq}R^2-2\gamma_\mathrm{eq}(\gamma_\mathrm{eq}-1)R+\gamma_\mathrm{eq}(\gamma_\mathrm{eq}-1)^2}{2R^2},
\end{align}
with $R_0\equiv2p/\sigma=2/(\eta(\gamma_\mathrm{eq}-1))$ normalized to $1$ by equivalently setting $\eta=2/(\gamma_\mathrm{eq}-1)$. 
%Under vanishing limit of NLO friction $\Delta p_\mathrm{NLO}\to0$, namely,  $\gamma_\mathrm{eq}\to\infty$, \eqref{eq:gammaeom1} reduces to \eqref{eq:gamma0}.

Following the above approximated approach, one can turn to the case with $\gamma^2$-scaling friction. The effective EOM is now modified as
\begin{align}\label{eq:REOM12}
\ddot{R}+2\frac{1-\dot{R}^2}{R}=\frac{\Delta p_\mathrm{dr}-\Delta p_\mathrm{LO}-\gamma^2\Delta p_{N\mathrm{LO}}}{\sigma}\left(1-\dot{R}^2\right)^\frac32,
\end{align}
or rewritten in terms of the Lorentz factor,
\begin{align}\label{eq:GEOM2}
\frac{\mathrm{d}\gamma}{\mathrm{d}R}+\frac{2\gamma}{R}=\eta(\gamma_\mathrm{eq}^2-\gamma^2),
\end{align}
where abbreviations  \cite{Ellis:2020nnr} 
\begin{align}
\eta\equiv\frac{\Delta p_\mathrm{NLO}}{\sigma},\quad \gamma_\mathrm{eq}\equiv\sqrt{\frac{\Delta p_\mathrm{dr}-\Delta p_\mathrm{LO}}{\Delta p_\mathrm{NLO}}}
\end{align}
are introduced for convenient parameterization. Directly solving \eqref{eq:GEOM2} gives rise to 
\begin{align}
\gamma(R)=-\frac{1}{\eta R}+\gamma_\mathrm{eq}\tanh(\gamma_\mathrm{eq}\eta R+iC),
\end{align}
which, after fixing the constant $C$ by the initial condition $\gamma(R_0)=1$, becomes
\begin{align}
\gamma(R)=-\frac{1}{\eta R}+\gamma_\mathrm{eq}\tanh\left(\gamma_\mathrm{eq}\eta(R-R_0)+\mathrm{arccoth}\left(\frac{\gamma_\mathrm{eq}\eta R_0}{1+\eta R_0}\right)\right),
\end{align}
namely,
\begin{align}\label{eq:gammaeom2}
\gamma(R)=\frac{1}{2R}-\frac{\gamma_\mathrm{eq}^2}{2R}+\gamma_\mathrm{eq}\tanh\left(\frac{2\gamma_\mathrm{eq}}{\gamma_\mathrm{eq}^2-1}(R-1)+\mathrm{arccoth}\left(\frac{2\gamma_\mathrm{eq}}{\gamma_\mathrm{eq}^2+1}\right)\right),
\end{align}
with $R$ normalized to $R_0=2/(\eta(\gamma_\mathrm{eq}^2-1))\equiv1$ by equivalently setting $\eta=2/(\gamma_\mathrm{eq}^2-1)$.  Unfortunately, this solution is not even real since the term of arccoth is always imaginary. 

Nevertheless, both \eqref{eq:gammaeom1} and \eqref{eq:gammaeom2}  admit a sensible limit at large $\gamma_\mathrm{eq}$,
\begin{align}
\lim\limits_{\gamma_\mathrm{eq}\to\infty}\gamma(R)=\frac23R+\frac{1}{3R^2},
\end{align}
as expected from the vanishing $\gamma$-dependent friction when the bubble wall exhibits runaway behavior as in the case without thermal friction. It also admits a sensible limit at large $R$,
\begin{align}
\lim\limits_{R\to\infty}\gamma(R)=\gamma_\mathrm{eq},
\end{align}
as expected from balancing the $\gamma$-independent and $\gamma$-dependent contributions to the total pressure changes when the bubble bubble wall ceases to accelerate at large $R$.

Before closing this section, one may wonder why it ends up with an imaginary solution \eqref{eq:gammaeom2} from the seemingly innocent effective EOM \eqref{eq:GEOM2}. This could be traced back to the normalization choice of $R_0=2/(\eta(h(\gamma_\mathrm{eq})-h(1)))$ for the friction term of form $\Delta p_\mathrm{fr}=-\Delta p_\mathrm{LO}-h(\gamma)\Delta p_{N\mathrm{LO}}$ with $h(\gamma)=\gamma^2$. Such a choice of $R_0$ should be obtained from the size of a critical bubble that stabilizes the total energy
\begin{align}
E&=4\pi R^2\sigma\gamma+\frac43\pi R^3V_\mathrm{eff}(\phi_-)-\frac43\pi R^3V_\mathrm{eff}(\phi_+)+\frac43\pi R^3|\Delta p_\mathrm{fr}|\\
&\equiv 4\pi R^2\sigma\gamma-\frac43\pi R^3(\Delta p_\mathrm{dr}+\Delta p_\mathrm{fr})\equiv 4\pi R^2\sigma\gamma-\frac43\pi R^3\sigma\eta(h(\gamma_\mathrm{eq})-h(\gamma)).\label{eq:Egeneral}
\end{align}
Here in the first line the first term is the Lorentz boost of the stored tension energy for a moving bubble wall, the second term subtracted with the third term is the potential energy with respect to some reference point set by the false vacuum, and the last term is a measure of the work done by the friction force on the bubble wall, which simply accounts for the energy exchanges between the bubble and plasma that should be included in the total energy. Hence there is an inconsistency between the definition of the bubble energy and the effective EOM \eqref{eq:GEOM2}. In fact, the corresponding Lagrangian from the Legendre transformation of the bubble energy does not lead to the effective EOM \eqref{eq:GEOM2}. As we will see shortly below, there is an additional correction term.

\subsubsection{Proposed effective description}\label{subsec:new}

The approximated estimation presented in section \ref{subsec:old}  is not physically self-consistent since the $\gamma$-dependent friction is added by hand into the EOM \eqref{eq:REOM0} that are derived from a Lagrangian without $\gamma$-dependence in its friction term in the first place. An appropriate question to ask is what kind of form for a Lagrangian to reproduce (1) the given $\gamma$-scaling friction term in its EOM, (2) the asymptotically flat limit at large bubble radius,  (3) the correct limit at vanishing $\gamma$-dependent  friction.

\begin{figure}
\centering
\includegraphics[width=0.49\textwidth]{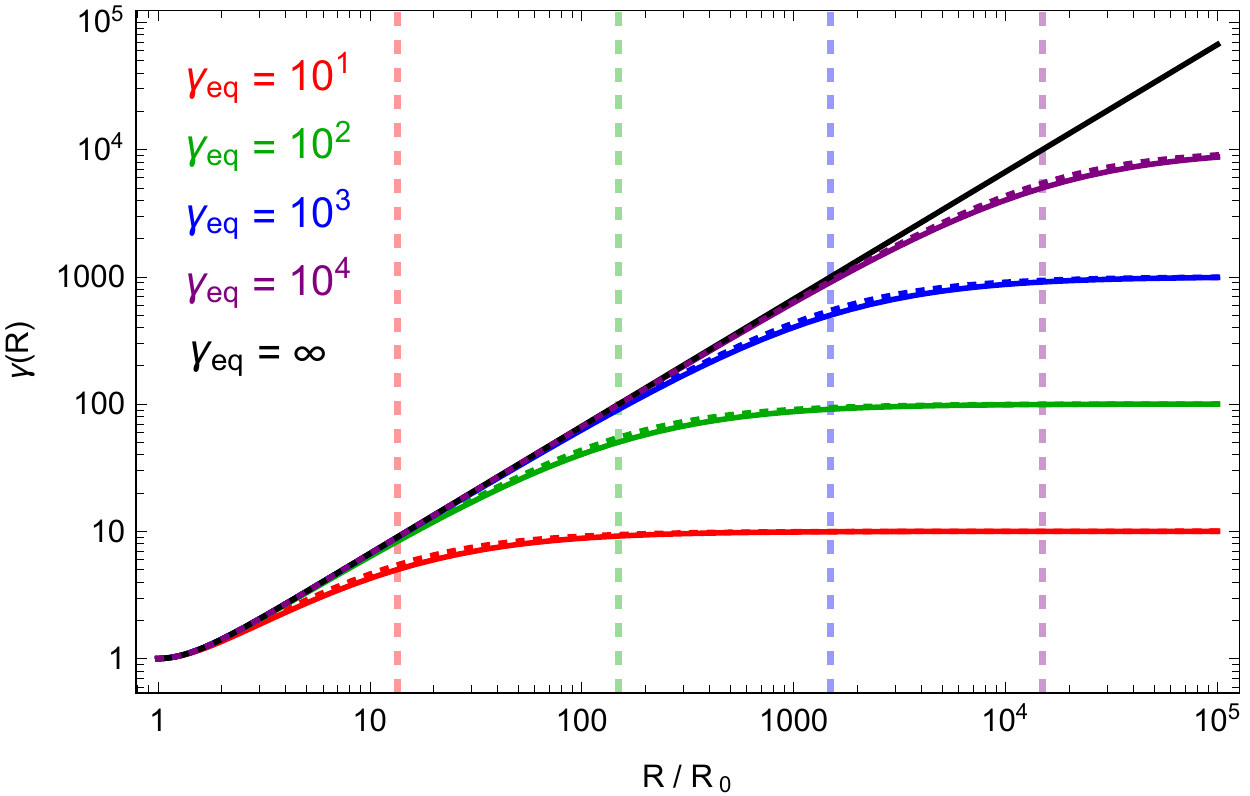}
\includegraphics[width=0.49\textwidth]{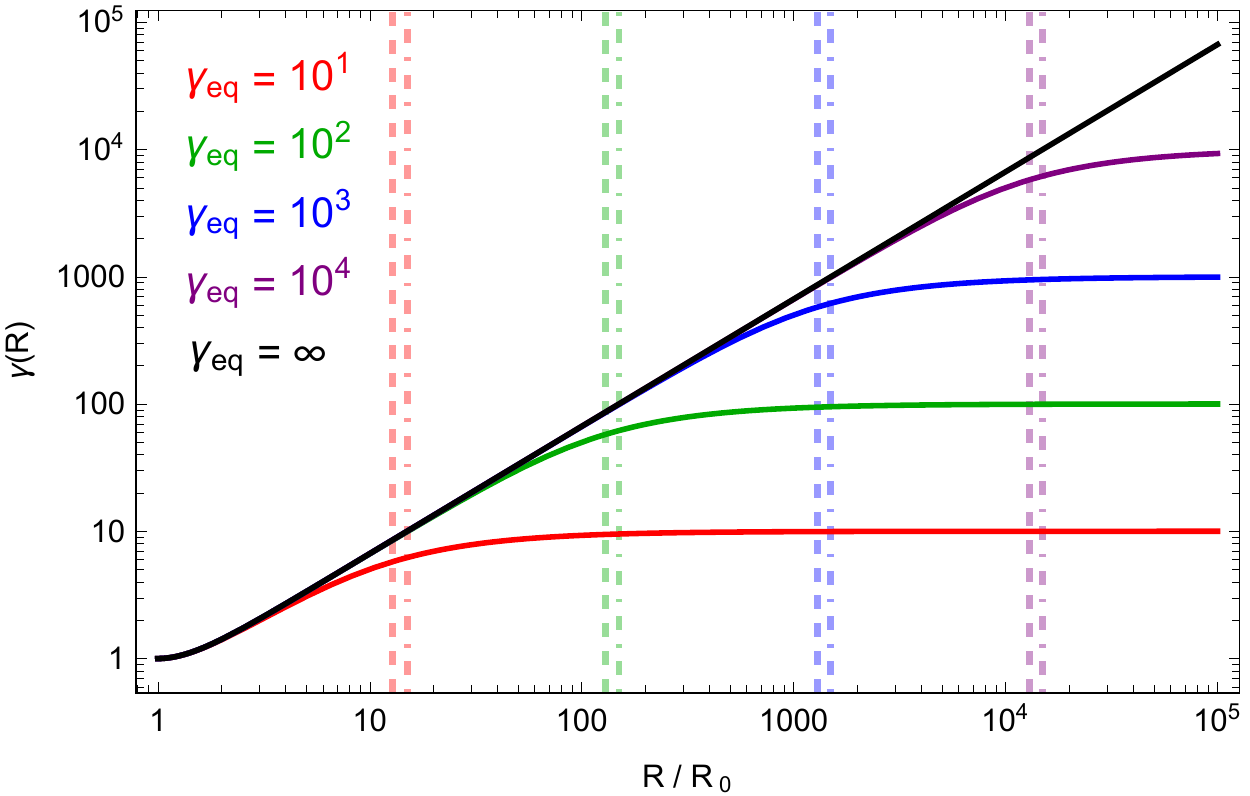}\\
\caption{Lorentz factor at given bubble radius during bubble expansion with $\gamma^1$-scaling friction (left) and $\gamma^2$-scaling friction (right) for some illustrative values of $\gamma_\mathrm{eq}$. The vertical dashed and dot-dashed lines are $R_\sigma$ and $R_\mathrm{eq}$, respectively. The dashed curves in the left panel are from a previous approximated solution \cite{Ellis:2019oqb}. }\label{fig:gammaR}
\end{figure}

We start with an general effective Lagrangian with its friction term separable in both $\dot{R}$ and $R$, namely,
\begin{align}\label{eq:Lguessgeneral}
L=-4\pi\sigma R^2\sqrt{1-\dot{R}^2}+\frac43\pi R^3\left(\Delta p_\mathrm{dr}-\Delta p_\mathrm{LO}+f(R)g(\dot{R})\right),
\end{align}
for some functions $f(R)$ and $g(\dot{R})$, of which the corresponding EOM reads
\begin{align}\label{eq:generaleom}
\left(1+\frac{f}{3\sigma}\frac{g''}{\gamma^3}R\right)\frac{\mathrm{d}\gamma}{\mathrm{d}R}+\frac{2\gamma}{R}=\frac{\Delta p_\mathrm{dr}-\Delta p_\mathrm{LO}}{\sigma}-\frac{3f+Rf'}{3\sigma}(\dot{R}g'-g),
\end{align}
with $f'\equiv\mathrm{d}f/\mathrm{d}R$ and $g'\equiv\mathrm{d}g/\mathrm{d}\dot{R}$ for short. To reproduce the friction $P_{1\to N}\equiv h(\gamma)\Delta p_{N\mathrm{LO}}$ in the EOM with arbitrary scaling function $h(\gamma)$, we have to solve following equations:
\begin{align}
3f(R)+Rf'(R)&=3\Delta p_{N\mathrm{LO}},\\
\dot{R}g'(\dot{R})-g(\dot{R})&=h(\gamma(\dot{R})),
\end{align}
of which the general solution should be of form
\begin{align}
f(R)&=\Delta p_{N\mathrm{LO}}+\frac{C_1}{R^3},\\
g(\dot{R})&=\dot{R}\left[C_2+\int_1^{\dot{R}}\frac{\mathrm{d}\dot{r}}{\dot{r}^2}h(\gamma(\dot{r}))\right].
\end{align}
Hereafter we will choose the integration constants $C_1=0$ for convenience while the presence of $C_2$ would not change our results. Putting the Lagrangian on-shell by
\begin{align}
L=-4\pi\sigma R^2\sqrt{1-\dot{R}^2}+\frac43\pi R^3\sigma\eta\left(h(\gamma_\mathrm{eq})+\dot{R}g'(\dot{R})-h(\gamma(\dot{R}))\right),
\end{align}
with abbreviations
\begin{align}\label{eq:abbreviation}
\eta\equiv\frac{\Delta p_{N\mathrm{LO}}}{\sigma},\quad h(\gamma_\mathrm{eq})\equiv\frac{\Delta p_\mathrm{dr}-\Delta p_\mathrm{LO}}{\Delta p_{N\mathrm{LO}}},
\end{align}
we would obtain the total energy of the bubble as
\begin{align}
E&=\frac{\partial L}{\partial\dot{R}}\dot{R}-L\nonumber\\
&=4\pi\sigma R^2\gamma+\frac43\pi R^3\sigma\eta(h(\gamma)-h(\gamma_\mathrm{eq}))+\frac43\pi R^3\sigma\eta\left(g''(\dot{R})-\gamma^3h'(\gamma)\right)\dot{R}^2\nonumber\\
&=4\pi\sigma R^2\gamma+\frac43\pi R^3\sigma\eta(h(\gamma)-h(\gamma_\mathrm{eq})),\label{eq:generalE}
\end{align}
where $g''(\dot{R})=\gamma^3h'(\gamma)$ is used to get the last line. 
%Therefore, the total energy of the bubble is simply the sum of the kinetic energy of the bubble wall and the net change of the total pressure. 
This is exactly the form of \eqref{eq:Egeneral}.
Directly solving \eqref{eq:generalE} for $\gamma(R)$ also yields a solution of the EOM \eqref{eq:generaleom} of form 
\begin{align}\label{eq:generalEOM}
\left(1+\frac{\eta R}{3}h'(\gamma)\right)\frac{\mathrm{d}\gamma}{\mathrm{d}R}+\frac{2\gamma}{R}=\eta(h(\gamma_\mathrm{eq})-h(\gamma))
\end{align}
if $E$ is treated as a constant. Therefore, our effective description is essentially the conservation law of the total energy $E$ of the bubble, which can be fixed as $E=4\pi\sigma R_0^2+\frac43\pi R_0^3\sigma\eta(h(1)-h(\gamma_\mathrm{eq}))$ by the initial static condition $\gamma(R=R_0)=1$ with the initial radius $R_0$ determined by $\mathrm{d}E/\mathrm{d}R|_{\gamma(R_0)=1}=0$ as
\begin{align}\label{eq:generalR0}
R_0=\frac{2}{\eta(h(\gamma_\mathrm{eq})-h(1))}.
\end{align}
Hence the solution of \eqref{eq:generalE} (or equivalently \eqref{eq:generalEOM}) reads,
\begin{align}
h(\gamma(R))+\frac{3\gamma(R)}{\eta R}=h(\gamma_\mathrm{eq})+\left(\frac{R_0}{R}\right)^3\left(\frac{3}{\eta R_0}+h(1)-h(\gamma_\mathrm{eq})\right),
\end{align}
which, after normalizing $R$ to $R_0\equiv1$ by equivalently setting $\eta=2/(h(\gamma_\mathrm{eq})-h(1))$, becomes
\begin{align}\label{eq:generalsol}
\frac{h(\gamma)-h(1)}{h(\gamma_\mathrm{eq})-h(1)}+\frac{3\gamma}{2R}=1+\frac{1}{2R^3}.
\end{align}
For $\gamma^1$-scaling friction  $P_{1\to 2}\equiv h(\gamma)\Delta p_{\mathrm{NLO}}$ with $h(\gamma)=\gamma$, $\gamma(R)$ is solved as
\begin{align}\label{eq:gammaR1}
\gamma(R)=\frac{2\gamma_\mathrm{eq}R^3+\gamma_\mathrm{eq}-1}{2R^3+3(\gamma_\mathrm{eq}-1)R^2}.
\end{align}
For $\gamma^2$-scaling friction $P_{1\to N}\equiv h(\gamma)\Delta p_{N\mathrm{LO}}$ with $h(\gamma)=\gamma^2$, $\gamma(R)$ is solved as
\begin{align}\label{eq:gammaR2}
\gamma(R)=\sqrt{\gamma_\mathrm{eq}^2+\frac{9(\gamma_\mathrm{eq}^2-1)^2}{16R^2}+\frac{\gamma_\mathrm{eq}^2-1}{2R^3}}-\frac{3(\gamma_\mathrm{eq}^2-1)}{4R}.
\end{align}
Both solutions admit following two limits,
\begin{align}
\lim\limits_{\gamma_\mathrm{eq}\to\infty}\gamma(R)&=\frac23R+\frac{1}{3R^2},\\
\lim\limits_{R\to\infty}\gamma(R)&=\gamma_\mathrm{eq}.
\end{align}
Therefore, we complete our effective description fulfilling the three conditions mentioned at the beginning of this section, and the solution \eqref{eq:gammaR1} and \eqref{eq:gammaR2} are shown as the solid curves in the left and right panels of Fig. \ref{fig:gammaR}, respectively, for some illustrative values of $\gamma_\mathrm{eq}$. Note that in the left panel, the approximated solution \eqref{eq:gammaeom1} is also shown as the dashed curves for comparison, although the normalization radius $R_0$ is not consistent with the effective EOM \eqref{eq:GEOM1}.

Before closing this section, it is worth noting that, there is an interesting phenomenon appeared in the coefficient of $\mathrm{d}\gamma/\mathrm{d}R$ of the EOM \eqref{eq:generalEOM}, namely,
\begin{align}
1+\frac{f}{3\sigma}\frac{g''}{\gamma^3}R=1+\frac{\eta}{3}h'(\gamma)R.
\end{align}
This suggests that in our effective description the surface tension of a thermal bubble out of thermal equilibrium receives a correction in the $\mathrm{d}\gamma/\mathrm{d}R$ term, which is proportional to both the bubble radius and the derivative of the thermal friction with respect to the Lorentz factor of the bubble wall velocity,
\begin{align}\label{eq:sigma}
\tilde{\sigma}=\sigma+\frac{R}{3}\frac{\mathrm{d}P_{1\to N}}{\mathrm{d}\gamma}.
\end{align}
This correction is negligible during the early stage of bubble expansion but starts dominating the tension growth when 
\begin{align}
R>R_\sigma=\frac{3}{\eta h'(\gamma(R_\sigma))}=\frac32\frac{h(\gamma_\mathrm{eq})-h(1)}{h'(\gamma(R_\sigma))}R_0,
\end{align}
after which the solution $\gamma(R)$ start to become asymptotically flat since the surface tension tends to shrink the bubble wall against expanding.  To see $R_\sigma$ quantitatively, we take two examples:
\begin{enumerate}
\item For $\gamma^1$-scaling friction $P_{1\to 2}\equiv h(\gamma)\Delta p_{\mathrm{NLO}}$ with $h(\gamma)=\gamma$, a roughly matching could be found as
\begin{align}
R_\sigma\equiv\frac32(\gamma_\mathrm{eq}-1)R_0\approx R_\mathrm{eq}\equiv\frac32\gamma_\mathrm{eq}R_0,
\end{align}
where $R_\mathrm{eq}$ proposed in \cite{Ellis:2019oqb} is estimated from the friction-free solution by
\begin{align}
\gamma(R)=\frac23\frac{R}{R_0}+\frac12\frac{R_0^2}{R^2}\approx\frac23\frac{R}{R_0}\equiv\gamma_\mathrm{eq}.
\end{align}
$R_\sigma$ is shown as the vertical dashed lines in the left panel.
\item For $\gamma^2$-scaling friction $P_{1\to N}\equiv h(\gamma)\Delta p_{N\mathrm{LO}}$ with $h(\gamma)=\gamma^2$, $R_\sigma$ is solved as a real positive (and larger than $R_0\equiv1$) root of 
\begin{align}
\gamma(R_\sigma)\frac{R_\sigma}{R_0}=\frac34(\gamma_\mathrm{eq}^2-1)\Leftrightarrow\gamma_\mathrm{eq}^2R_\sigma^3-27(\gamma_\mathrm{eq}^2-1)^2R_\sigma+8(\gamma_\mathrm{eq}^2-1)=0,
\end{align}
which is shown as the vertical dashed lines in the right panel of Fig. \ref{fig:gammaR} for given $\gamma_\mathrm{eq}$. Note that in \cite{Ellis:2020nnr} $R_\mathrm{eq}$ from the presence of $\gamma^2$-scaling friction is still  approximated by $\frac32\gamma_\mathrm{eq}R_0$ also shown as the dot-dashed lines in the right panel of Fig. \ref{fig:gammaR}, which is close to our $R_\sigma$ by amount of 
\begin{align}
\frac{R_\sigma\gamma_\sigma}{R_\mathrm{eq}\gamma_\mathrm{eq}}=\frac12(1-\gamma_\mathrm{eq}^{-2}).
\end{align}
\end{enumerate}

\section{Efficiency factor for bubble collisions}\label{sec:efficiency}

The prediction of the GW background from the cosmic first-order phase transition heavily relies on the so-called bubble collision efficiency factor   \cite{Ellis:2019oqb,Ellis:2020nnr}
\begin{align}
\kappa_\mathrm{col}&=\frac{E_\mathrm{wall}(R_\mathrm{col})}{\frac43\pi R_\mathrm{col}^3|\Delta V_\mathrm{eff}|}
\end{align}
upon bubble collisions time with mean bubble radius $R_\mathrm{col}$, where the energy fraction stored in the expanding bubble wall is calculated from
\begin{align}
E_\mathrm{wall}(R_\mathrm{col})&=4\pi R_\mathrm{col}^2\int_{R_0}^{R_\mathrm{col}}\frac{\mathrm{d}R}{3}\bigg[|\Delta V_\mathrm{eff}|-\Delta p_\mathrm{LO}-h(\gamma(R))\Delta p_{N\mathrm{LO}}\bigg].
\end{align}
We start with reviewing the estimations of \cite{Ellis:2020nnr} for $h(\gamma)=\gamma^2$ with approximation $R_\mathrm{col}\gg R_0$:
\begin{enumerate}
\item $R_\mathrm{col}<R_\mathrm{eq}$ : In this regime with accelerating bubble wall, we could approximate $\gamma(R)\approx\frac23R$ so that
\begin{align}
E_\mathrm{wall}(R_\mathrm{col})&\approx4\pi R_\mathrm{col}^2\int_0^{R_\mathrm{col}}\frac{\mathrm{d}R}{3}\left(|\Delta V_\mathrm{eff}|-\Delta p_\mathrm{LO}-\frac49R^2\Delta p_{N\mathrm{LO}}\right)\nonumber\\
&=\frac43\pi R_\mathrm{col}^3(|\Delta V_\mathrm{eff}|-\Delta p_\mathrm{LO})-\frac43\pi R_\mathrm{col}^3\gamma_\mathrm{col}^2\frac{\Delta p_{N\mathrm{LO}}}{3},
\end{align}
with $\gamma_\mathrm{col}\approx\frac23R_\mathrm{col}$, leading to the bubble collision efficiency factor as
\begin{align}
\kappa_\mathrm{col}&=\left(1-\frac{\Delta p_\mathrm{LO}}{|\Delta V_\mathrm{eff}|}\right)-\frac{\gamma_\mathrm{col}^2}{3}\frac{\Delta p_{N\mathrm{LO}}}{|\Delta V_\mathrm{eff}|}\nonumber\\
&=\left(1-\frac13\frac{\gamma_\mathrm{col}^2}{\gamma_\mathrm{eq}^2}\right)\left(1-\frac{\alpha_\infty}{\alpha}\right),\label{eq:kappacol21}
\end{align}
where in the second line some conventional abbreviations are adopted,
\begin{align}
\alpha_\infty\equiv\frac{\Delta p_\mathrm{LO}}{\rho_R}, \, 
\alpha\equiv\frac{|\Delta V_\mathrm{eff}|}{\rho_R}, \,
\gamma_\mathrm{eq}\equiv\sqrt{\frac{|\Delta V_\mathrm{eff}|-\Delta p_\mathrm{LO}}{\Delta p_{N\mathrm{LO}}}}=\sqrt{\left(1-\frac{\alpha_\infty}{\alpha}\right)\bigg/\frac{\Delta p_{N\mathrm{LO}}}{|\Delta V_\mathrm{eff}|}}
\end{align}
with $\rho_R$ the background radiation energy density at bubble collision time.
\item $R_\mathrm{col}>R_\mathrm{eq}$ : In this regime with the balance between $|\Delta V_\mathrm{eff}|-\Delta p_\mathrm{LO}$ and $\gamma_\mathrm{eq}^2\Delta p_{N\mathrm{LO}}$ , the integrand in $E_\mathrm{wall}$ is approximately zero over the interval $[R_\mathrm{eq}, R_\mathrm{col}]$ so that
\begin{align}
E_\mathrm{wall}(R_\mathrm{col})&\approx4\pi R_\mathrm{col}^2\int_0^{R_\mathrm{eq}}\frac{\mathrm{d}R}{3}\left(|\Delta V_\mathrm{eff}|-\Delta p_\mathrm{LO}-\frac49R^2\Delta p_{N\mathrm{LO}}\right)\nonumber\\
&=\frac43\pi R_\mathrm{col}^2R_\mathrm{eq}(|\Delta V_\mathrm{eff}|-\Delta p_\mathrm{LO})-\frac43\pi R_\mathrm{col}^2R_\mathrm{eq}\gamma_\mathrm{eq}^2\frac{\Delta p_{N\mathrm{LO}}}{3},
\end{align}
with $\gamma_\mathrm{eq}\approx\frac23R_\mathrm{eq}$, leading to the bubble collision efficiency factor as
\begin{align}
\kappa_\mathrm{col}&=\frac{R_\mathrm{eq}}{R_\mathrm{col}}\left(1-\frac{\Delta p_\mathrm{LO}}{|\Delta V_\mathrm{eff}|}\right)-\frac{R_\mathrm{eq}}{R_\mathrm{col}}\frac{\gamma_\mathrm{eq}^2}{3}\frac{\Delta p_{N\mathrm{LO}}}{|\Delta V_\mathrm{eff}|}\nonumber\\
&=\frac23\frac{\gamma_\mathrm{eq}}{\gamma_\mathrm{col}}\left(1-\frac{\alpha_\infty}{\alpha}\right),\label{eq:kappacol22}
\end{align}
where $\gamma_\mathrm{col}\approx\frac23R_\mathrm{col}$ is the would-be friction-free Lorentz factor despite the fact that $\gamma(R_\mathrm{col})\neq\gamma_\mathrm{col}$ since $R_\mathrm{col}>R_\mathrm{eq}$.
\end{enumerate}
Both estimations \eqref{eq:kappacol21} and \eqref{eq:kappacol22} for the bubble collision efficiency factor are functions of $R_\mathrm{col}$ for given $\alpha_\infty/\alpha$ and $\gamma_\mathrm{eq}$ and valid only for the large radius $R_\mathrm{col}\gg R_0$ at bubble collisions.

\begin{figure}
\centering
\includegraphics[width=0.49\textwidth]{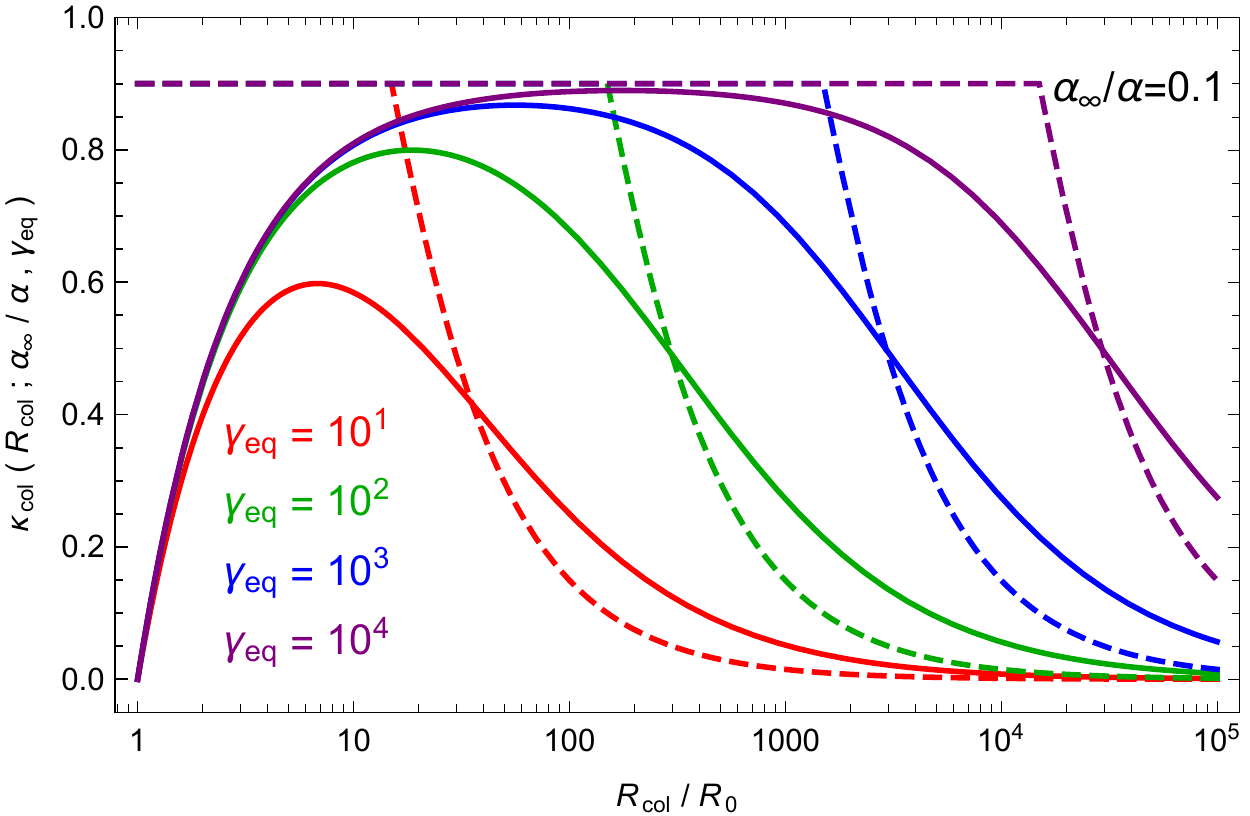}
\includegraphics[width=0.49\textwidth]{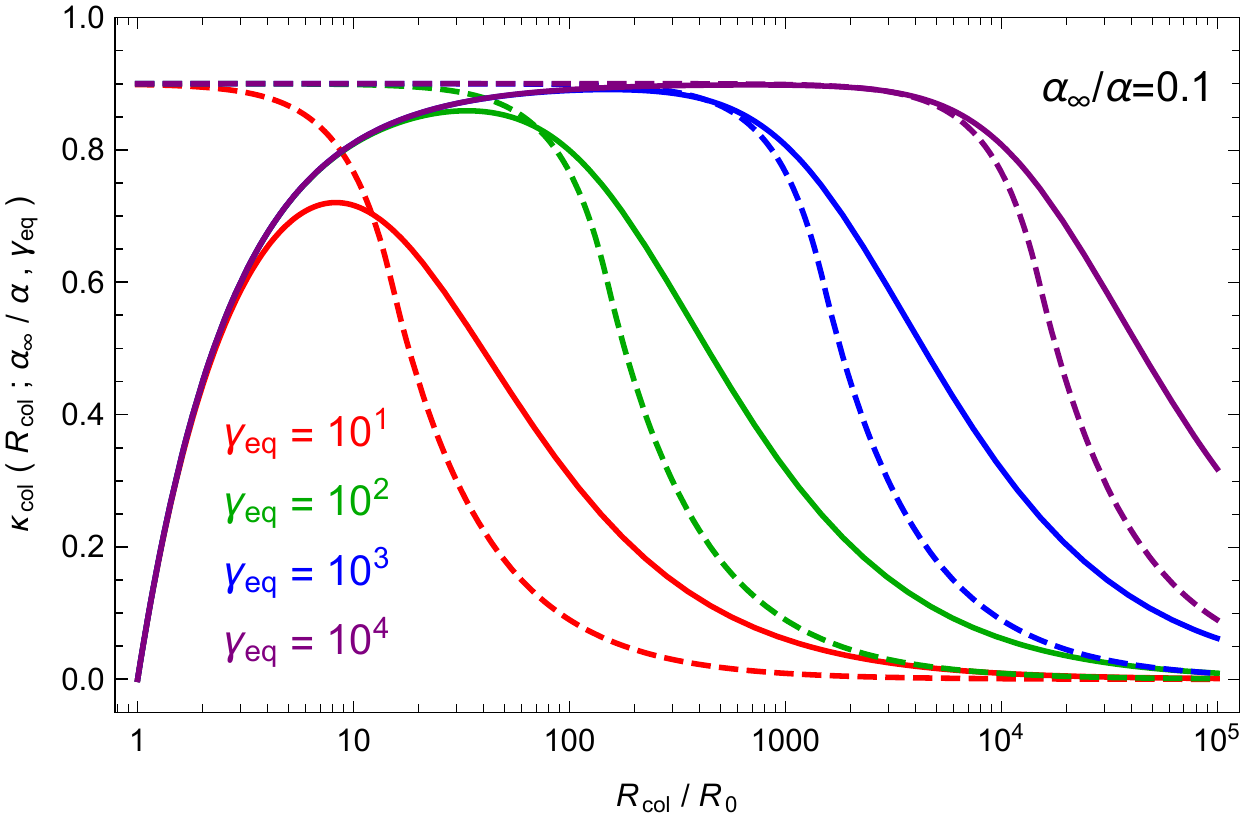}\\
\caption{The bubble collision efficiency factor $\kappa_\mathrm{col}$ at input bubble collision radius $R_\mathrm{col}$ for a given $\alpha_\infty/\alpha=0.1$ and some illustrative values of $\gamma_\mathrm{eq}$ from our computation (solid curves) compared to the previous estimation (dashed curves) with the $\gamma^1$-scaling friction (left) and $\gamma^2$-scaling friction (right). }\label{fig:kappacol}
\end{figure}

However, since we already have the full evolution solution \eqref{eq:gammaR2}, we could directly compute the bubble collision efficiency factor as
\begin{align}
\kappa_\mathrm{col}&=\frac{\frac43\pi R_\mathrm{col}^2}{\frac43\pi R_\mathrm{col}^3}\int_{R_0\equiv1}^{R_\mathrm{col}}\mathrm{d}R\left[\frac{|\Delta V_\mathrm{eff}|-\Delta p_\mathrm{LO}}{|\Delta V_\mathrm{eff}|}-\gamma(R)^2\frac{\Delta p_{N\mathrm{LO}}}{|\Delta V_\mathrm{eff}|}\right]\nonumber\\
&=\frac{1}{R_\mathrm{col}}\int_1^{R_\mathrm{col}}\mathrm{d}R\left[\left(1-\frac{\alpha_{\infty}}{\alpha}\right)-\left(1-\frac{\alpha_{\infty}}{\alpha}\right)\frac{\gamma(R)^2}{\gamma_\mathrm{eq}^2}\right]\nonumber\\
&=\left(1-\frac{\alpha_{\infty}}{\alpha}\right)\int_{R_\mathrm{col}^{-1}}^1\mathrm{d}\left(\frac{R}{R_\mathrm{col}}\right)\left[1-\frac{\gamma(R)^2}{\gamma_\mathrm{eq}^2}\right],\label{eq:kappacol2}
\end{align}
which is also a function of $R_\mathrm{col}$ for given $\alpha_\infty/\alpha$ and $\gamma_\mathrm{eq}$ and valid for all $R_\mathrm{col}$. In the right panel of Fig. \ref{fig:kappacol}, we plot our result \eqref{eq:kappacol2} (solid curves) compared to the previous estimations \eqref{eq:kappacol21} and \eqref{eq:kappacol22} (dashed cures) for given $\alpha_\infty/\alpha=0.1$ and some illustrative values of $\gamma_\mathrm{eq}$. As seen from the solid curves, the bubble collision efficiency factor is growing until that the bubble wall starts approaching the terminal velocity, after which the bubble collision efficiency factor starts decreasing. This meets the naive expectation that most of the released vacuum energy goes into accelerating the bubble wall before starting approaching the terminal velocity, after which most of released vacuum energy dissipates into the plasma. As for the dashed curves, despite of the small collision radius regime where the previous estimations  \eqref{eq:kappacol21} and \eqref{eq:kappacol22} simply cannot be applied, it also underestimates the bubble collision efficiency factor in the regime where the bubble collision radius is around one order of  magnitude larger than the radius when the bubble wall starts approaching the terminal velocity, namely, $R_\mathrm{eq}\lesssim R_\mathrm{col}\lesssim\mathcal{O}(100)R_\mathrm{eq}$, which would lead to larger GW signals from bubble collisions due to its quadratic dependence on the bubble collision efficiency factor.  Nevertheless, these two estimations marginally agree with each other in the large collision radius regime but with a small terminal velocity, namely $R_\mathrm{col}\gg\mathcal{O}(100)R_\mathrm{eq}$. Therefore, our result would be important for the cases when bubbles collide at a radius $R_\mathrm{col}/R_0\lesssim\mathcal{O}(100)\gamma_\mathrm{eq}$, which could be relevant even for the late-time stage of expansion for a sufficiently large $\gamma_\mathrm{eq}$ needed for the large GWs signals from the first-order phase transition.

Since the calculation for the thermal friction on the $\gamma$-dependence is not settled down yet, we also present here the results of bubble collision efficiency factor with $\gamma^1$-scaling friction. With the solved solution \eqref{eq:gammaR1} and the abbreviations
\begin{align}
\alpha_\infty\equiv\frac{\Delta p_\mathrm{LO}}{\rho_R}, \, 
\alpha\equiv\frac{|\Delta V_\mathrm{eff}|}{\rho_R}, \,
\gamma_\mathrm{eq}\equiv\frac{|\Delta V_\mathrm{eff}|-\Delta p_\mathrm{LO}}{\Delta p_{\mathrm{NLO}}}=\left(1-\frac{\alpha_\infty}{\alpha}\right)\bigg/\frac{\Delta p_{\mathrm{NLO}}}{|\Delta V_\mathrm{eff}|},
\end{align}
the efficiency factor for the bubble collisions could be computed as
\begin{align}
\kappa_\mathrm{col}&=\frac{\frac43\pi R_\mathrm{col}^2}{\frac43\pi R_\mathrm{col}^3}\int_{R_0\equiv1}^{R_\mathrm{col}}\mathrm{d}R\left[\frac{|\Delta V_\mathrm{eff}|-\Delta p_\mathrm{LO}}{|\Delta V_\mathrm{eff}|}-\gamma(R)\frac{\Delta p_{\mathrm{NLO}}}{|\Delta V_\mathrm{eff}|}\right]\nonumber\\
&=\frac{1}{R_\mathrm{col}}\int_1^{R_\mathrm{col}}\mathrm{d}R\left[\left(1-\frac{\alpha_{\infty}}{\alpha}\right)-\left(1-\frac{\alpha_{\infty}}{\alpha}\right)\frac{\gamma(R)}{\gamma_\mathrm{eq}}\right]\nonumber\\
&=\left(1-\frac{\alpha_{\infty}}{\alpha}\right)\int_{R_\mathrm{col}^{-1}}^1\mathrm{d}\left(\frac{R}{R_\mathrm{col}}\right)\left[1-\frac{\gamma(R)}{\gamma_\mathrm{eq}}\right]\nonumber\\
&=\frac{1-\alpha_\infty/\alpha}{27\gamma_\mathrm{col}^2\gamma_\mathrm{eq}(\gamma_\mathrm{eq}-1)}\left[\gamma_\mathrm{col}(3\gamma_\mathrm{eq}-4)(3\gamma_\mathrm{eq}-1)^2\log\frac{3(\gamma_\mathrm{eq}+\gamma_\mathrm{col}-1)}{3\gamma_\mathrm{eq}-1}\right.\nonumber\\
&\qquad\qquad\qquad\qquad\qquad\left.-4\gamma_\mathrm{col}\log\frac{2}{3\gamma_\mathrm{col}}-2(\gamma_\mathrm{eq}-1)(3\gamma_\mathrm{col}-2)\right],\label{eq:kappacol1}
\end{align}
with $\gamma_\mathrm{col}\approx\frac23R_\mathrm{col}$  the would-be friction-free Lorentz factor. As a comparison, the bubble collision efficiency factor in \cite{Ellis:2019oqb} was estimated as
\begin{align}\label{eq:kappacol0}
\kappa_\mathrm{col}=\begin{cases}
1-\frac{\alpha_\infty}{\alpha}, & \gamma_\mathrm{col}<\gamma_\mathrm{eq},\\
\frac{\gamma_\mathrm{eq}}{\gamma_\mathrm{col}}\left[1-\frac{\alpha_\infty}{\alpha}\left(\frac{\gamma_\mathrm{eq}}{\gamma_\mathrm{col}}\right)^2\right], & \gamma_\mathrm{col}>\gamma_\mathrm{eq}.
\end{cases}
\end{align}
Note that \eqref{eq:kappacol0} was estimated in a different ways in \cite{Ellis:2019oqb} from the improved estimations \eqref{eq:kappacol21} and \eqref{eq:kappacol22} in \cite{Ellis:2020nnr}. 
We plot our result \eqref{eq:kappacol1} (solid curves) and previous estimation \eqref{eq:kappacol0} (dashed curves) in the left panel of Fig. \ref{fig:gammaR} for given $\alpha_\infty/\alpha=0.1$ and some illustrative values of $\gamma_\mathrm{eq}$. The discussion goes parallel as in the case with $\gamma^2$-scaling friction. It is easy to see that our effective description for calculating the efficiency factor of bubble collisions could be generalized into any $\gamma$-scaling function $h(\gamma)$ by
\begin{align}
\kappa_\mathrm{col}=\left(1-\frac{\alpha_{\infty}}{\alpha}\right)\int_{R_\mathrm{col}^{-1}}^1\mathrm{d}\left(\frac{R}{R_\mathrm{col}}\right)\left[1-\frac{h(\gamma(R))}{h(\gamma_\mathrm{eq})}\right]
\end{align}
with $\gamma(R)$ solved from \eqref{eq:generalsol} and $\gamma_\mathrm{eq}$ defined by \eqref{eq:abbreviation}.

\section{Conclusions and discussions}\label{sec:conclusion}

In this paper, we revisit the bubble expansion issue but with the inclusion of the thermal friction term of arbitrary $\gamma$-scaling, and the resulted effective EOM is derived for the first time as
\begin{align}
\left(\sigma+\frac{R}{3}\frac{\mathrm{d}|\Delta p_\mathrm{fr}|}{\mathrm{d}\gamma}\right)\frac{\mathrm{d}\gamma}{\mathrm{d}R}+\frac{2\sigma\gamma}{R}=|\Delta V_\mathrm{eff}|-|\Delta p_\mathrm{fr}|.
\end{align}
The bubble collision efficiency factor is re-calculated by the virtue of an effective description of the bubble expansion that can consistently reproduces the arbitrary $\gamma$-scaling friction in the effective EOM of bubble wall expansion. The resulted bubble collision efficiency factor in the case of $\gamma^2$-scaling friction is larger than a recently updated estimation, leading to a larger GW signal from bubble collisions even though the inclusion of $\gamma^2$-scaling friction has already suppressed the contribution from bubble collisions compared to the $\gamma^1$-scaling friction. Several comments are given below:

Firstly, our effective description to reproduce the given $\gamma$-scaling friction term in the effective EOM \eqref{eq:generalEOM} of bubble wall expansion is essentially the conservation law of the total energy of an expanding bubble, which is valid for any $\gamma$-dependent friction term with arbitrary $\gamma$-scaling function input by specific microscopic models of particle physics.

Secondly, a byproduct of our effective description is the presence of a scale-dependent correction \eqref{eq:sigma} to the surface tension of the cosmic bubble expansion in the plasma out of thermal equilibrium, which merits further numerical study and laboratory test if we can measure the surface tension experimentally in the cold-atom analog \cite{Fialko_2015,Fialko:2016ggg,Braden:2017add,Billam:2018pvp,Braden:2019vsw,Braden:2018tky} of vacuum decay but with thermal friction.

Thirdly, we also suggest a new form of parameterization for the out-of-equilibrium term as $-\eta_T(u^\mu\partial_\mu\phi)^2$, which can reproduce the $(\gamma^2-1)$-scaling of the friction term \eqref{eq:newfriction}. It would be interesting to include this parameterized friction term in the numerical simulations of bubble expansion in future study to see if our effective picture of bubble expansion could be reproduced.

\acknowledgments
We thank Jiang-Hao Yu for helpful comments on the manuscript.
RGC is supported by the National Natural Science Foundation of China Grants No. 11947302, No. 11991052, No. 11690022, No. 11821505 and No. 11851302, and by the Strategic Priority Research Program of CAS Grant No. XDB23030100, and by the Key Research Program of Frontier Sciences of CAS.
SJW was supported by the postdoctoral scholarship of Tufts University from NSF when part of this work was done at Tufts University. 

%\appendix

\bibliographystyle{JHEP}
\bibliography{ref}

\end{document}